\documentclass[useAMS]{mn2e}
\usepackage{graphicx}
\usepackage{txfonts}
\usepackage{ctable}
\usepackage{threeparttable}

\title[the dusty progenitors of the MW]{Decoding the stellar fossils of the dusty Milky Way progenitors}
\author[de Bennassuti et al.]{Matteo de Bennassuti$^{1, 2}$\thanks{E-mail:
matteo.debennassuti@oa-roma.inaf.it}, Raffaella Schneider$^{2}$, Rosa Valiante$^{2}$,
Stefania Salvadori$^{3}$\\
$^{1}$Dipartimento di Fisica, Sapienza, Universit$\grave{a}$ di Roma, Piazzale Aldo Moro 5, 00185, Roma, Italy\\
$^{2}$INAF - Osservatorio Astronomico di Roma, Via di Frascati 33, 00040 Monte Porzio Catone, Italy\\
$^{3}$Kapteyn Astronomical Institute, University of Groningen, Landleven 12, 9747 AD Groningen, The Netherlands\\
}
\begin{document}

\label{firstpage}

\date{Accepted . Received}

\pagerange{\pageref{firstpage}--\pageref{lastpage}} 

\pubyear{2014}

\maketitle

\begin{abstract}
We investigate the metallicity distribution function (MDF) in the Galactic halo and the relative fraction of Carbon-normal and Carbon-rich stars. To this aim, we use an improved version of the semi-analytical code GA\textsc{laxy} ME\textsc{rger} T\textsc{ree and} E\textsc{volution} (\textsc{gamete}), that reconstructs the hierarchical merger tree of the MW, following the star formation history and the metal and dust evolution in individual progenitors. The predicted scaling relations between the dust, metal and gas masses for MW progenitors show a good agreement with observational data of local galaxies and of Gamma Ray Burst (GRB) host galaxies at $\rm 0.1 < z < 6.3$. 
Comparing the simulated and the observed MDF, we find that in order to predict the formation of hyper-iron-poor stars at $\rm [Fe/H] < -4$, faint SN explosions have to dominate the metal yields produced by Pop III stars, disfavoring a Pop III IMF that extends  to stellar masses $\rm > 140\, M_{\odot}$, into the Pair-Instability SN progenitor mass range. The relative contribution of C-normal and C-enhanced stars to the MDF and its dependence on [Fe/H] points to a scenario where the Pop III/II transition is driven by dust-cooling and the first low-mass stars form when the dust-to-gas ratio in their parent clouds exceeds a critical value of ${\cal D}_{\rm crit} = 4.4 \times 10^{-9}$. Other transition criteria do not predict any C-normal stars below [Fe/H] $<$ -4, at odds with observations.

\end{abstract}


\section{Introduction}
\label{sec:intro}
The physical conditions which enable the formation of the first
low-mass and long-lived stars to form in the Universe is still
a subject of debate. According to the most recent numerical studies
of primordial star formation in the first mini-halos, the final
outcome appears to be the formation of a massive stellar binary or
of a small multiple of massive stars (for a recent comprehensive 
review see Bromm 2013). Although this represents a paradigm shift 
compared to the original ``standard model for the first stars",
which presumed the first Pop III stars to be very massive and to 
form in isolation, we are still far from being able to predict
the resulting stellar mass spectrum. The best available indications
from numerical studies suggest that the first stars formed in a 
broad mass range, from 10 to 1000\,  $\rm M_\odot$,  but that most
of them are distributed around a few tens to a few hundred solar masses
(Hosokawa et al. 2011; Greif et al. 2012; Turk et al. 2012; Hirano et al. 2014). 
Occasionally, one of the
secondary protostars formed during disk fragmentation is seen to migrate
to higher stellar orbit and may eventually form a low-mass primordial 
stars. Yet, no evidence for the existence of such primordial low-mass
stars has been found in surveys of the Galactic Halo and in nearby dwarfs
(Beers \& Christlieb 2005; Tolstoy et al. 2009).   

Once the first supernovae explode and start to seed the gas 
with metals and dust grains, the physical properties of 
star forming regions change, eventually allowing the formation
of the first low-mass Pop II stars. Gas at low metallicity can 
achieve larger cooling rates through additional molecular 
species (HD, OH, CO, H$_2$O), fine structure line--cooling (mostly OI
and CII), and thermal emission from dust grains (Omukai 2000; 
Schneider et al. 2002; Omukai et al. 2005). 
The relative importance of these coolants depends on the density (time)
regime during the collapse and on the initial metallicity and dust 
content of the collapsing core. 

Cooling due to line emission affects the thermal evolution of 
collapsing cores at low-to-intermediate densities: when 
the total metallicity of the gas is $\rm Z < 10^{-2} Z_{\odot}$,
molecular cooling dominates at $\rm n < 10^5 cm^{-3}$, 
with an efficiency which increases with metallicity because of 
the larger molecular formation rate on the surface of dust grains 
(Omukai et al. 2005; Schneider et al. 2006, 2012a). For more metal  
enriched clouds, $\rm Z \ge 10^{-2} Z_{\odot}$,  OI and CII 
fine-structure line cooling dominate the thermal evolution at
$\rm n < 10^{4} cm^{-3}$, until the NLTE-LTE transition occurs
and the cooling efficiency decreases. Bromm \& Loeb (2003) have
derived the critical O and C abundances required to overcome
the compressional heating rate and fragment the gas, finding $\rm [C/H]_{\rm cr} = -3.5$
and $\rm [O/H]_{\rm cr} = -3.05$. 
Frebel, Johnson \& Bromm (2007) have further 
explored the role of metal fine-structure line cooling by introducing the so-called transition discriminant -
 a combination of the C and O abundances -  that enables the formation of Pop II stars when its value exceeds a critical
threshold of $\rm D_{trans,cr} = -3.5 \pm 0.2$. Comparing this theoretical prediction with observations of metal-poor stars in 
the halo of the Milky Way (MW) and its dwarf satellites, they find that most of the stars have $\rm D_{trans} >  D_{trans,cr}$, with 
the exception of SDSS J102915+172927, a star with $\rm [Fe/H] = -4.99$ that falls in the forbidden zone for metal fine-structure
line cooling (Caffau et al. 2011b). Note also that given the typical gas temperatures and
densities where line-induced fragmentation takes place, the associated Jeans--unstable
fragment masses are still relatively large, with masses $\rm \ge 10 M_{\odot}$ (Schneider et al. 2006).
In a recent numerical simulation, Safranek-Shrader, Milosavljevi{\'c} \& Bromm (2014)
have confirmed that there is a clear metallicity threshold for widespread
fragmentation between $[10^{-4} - 10^{-3}] Z_{\odot}$ in high-redshift, 
atomic-cooling halos: fine-structure line cooling allows the gas to cool to the CMB
temperatures and form a cluster of size $\sim 1$~pc, with 
typical fragment masses $\sim 50 - 100 \rm \,M_{\odot}$, consistent with the Jeans masses 
expected at the end of the fragmentation phase. 
Although the simulations do not follow the evolution of the cluster to higher densities,
taking these results at face value it appears that fine-structure line cooling is not
capable of producing solar-mass fragments that have the potential to survive until the
present-day, corroborating previous findings based on semi-analytical models (Schneider
et al. 2006).

A pervasive fragmentation mode that allows the formation of solar or sub-solar mass 
fragments is activated at higher gas densities, $\rm n > 10^{10}-10^{12} cm^{-3}$, when
dust grains are collisionally excited and emit continuum radiation, thereby decreasing
the gas temperature until it becomes thermally coupled with the dust temperature 
$\rm T = T_{dust}$ or the gas becomes optically thick (Schneider et al. 2002).  
The efficiency of this physical process depends on the dust-to-gas ratio and on the 
total grain cross-section, with smaller grains providing a larger 
contribution to cooling (Schneider et al. 2006). Hence, the minimal conditions for 
dust-induced fragmentation have been expressed in terms of a minimal critical dust-to-gas
ratio, 
\begin{equation}
S {\cal D_{\rm cr}} = 1.4 \times 10^{-3} \ {\rm cm^2 g^{-1}} \left( \frac{T}{10^3 \ {\rm K}} \right) ^{-1/2}
\left( \frac{n}{10^{12} \rm cm^{-3}} \right)^{-1/2},
\label{eq:SD_crit}
\end{equation}
where $S$ is the total cross-section of dust grains per unit dust mass and the expression 
holds in the regime where dust cooling is effective, hence $\rm T_{dust} << T$. 
An extensive parameter-space exploration shows that ${\cal D_{\rm cr}} \sim  4.4 \times 10^{-9}$
can be considered as a good representative value for the minimal dust enrichment required
to activate dust-induced fragmentation (Schneider et al. 2012a). These results have been confirmed by
numerical simulations (Tsuribe \& Omukai 2006, 2008; Clark et al. 2008; Omukai et al. 2010; Dopcke et al. 2011). 
In addition, the observed properties of SDSS J102915+172927 provide a strong indication that low-mass star formation
may have occurred in the absence of efficient metal-line cooling, supporting a dust-driven transition 
(Schneider et al. 2014; Chiaki et al. 2014).

%%%%%%%%%%%%%%%%%%%%%%%%%%%%%%%%%%%%%%%%%%%%%%%%%%%%%%%%%
%%% Figure comparison Analytical - Simulated EPS
\begin{figure*}
\hspace{-0.4cm}
\includegraphics [width=5.8cm]{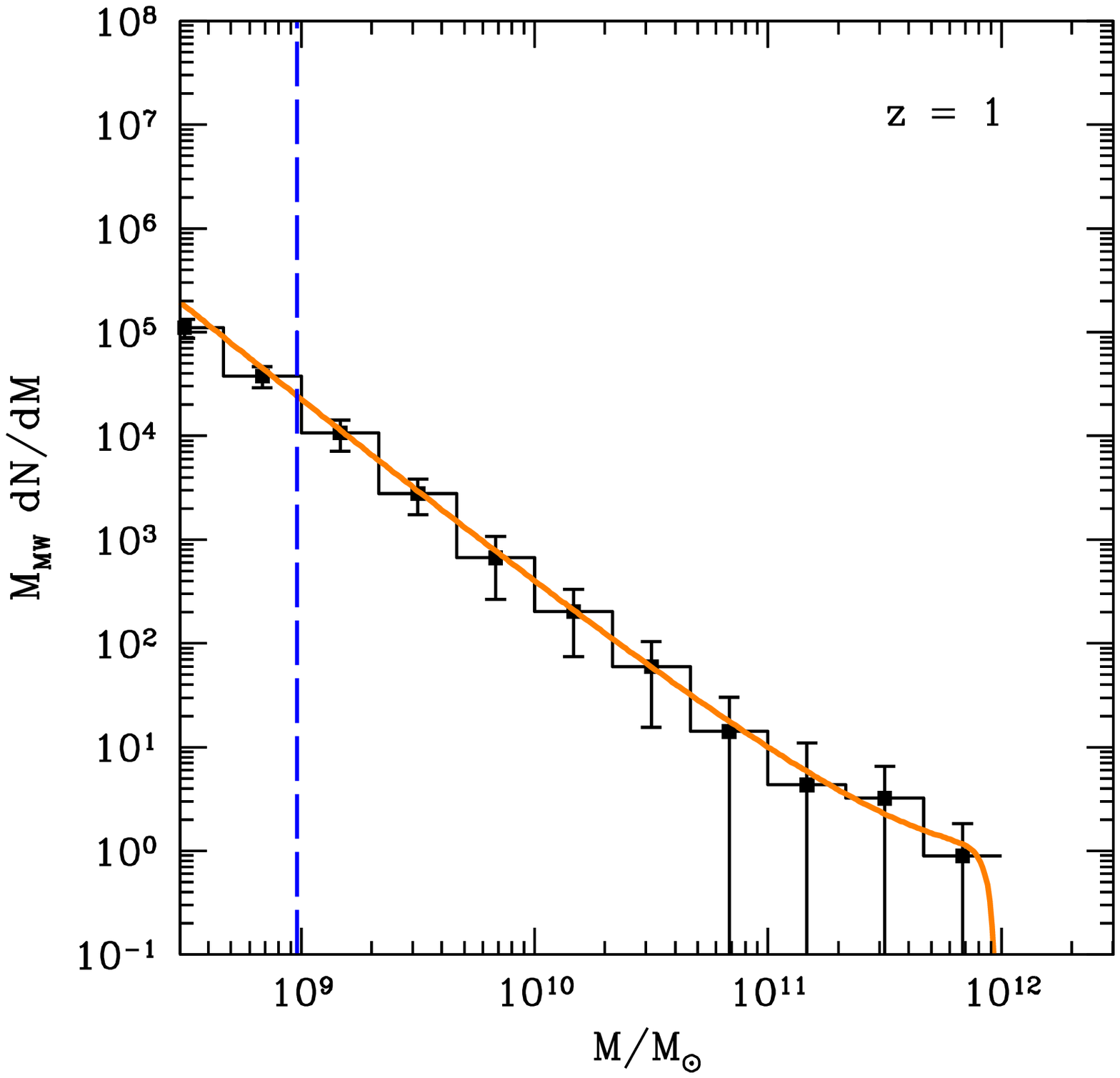}
\hspace{-1.87cm}
\includegraphics [width=5.8cm]{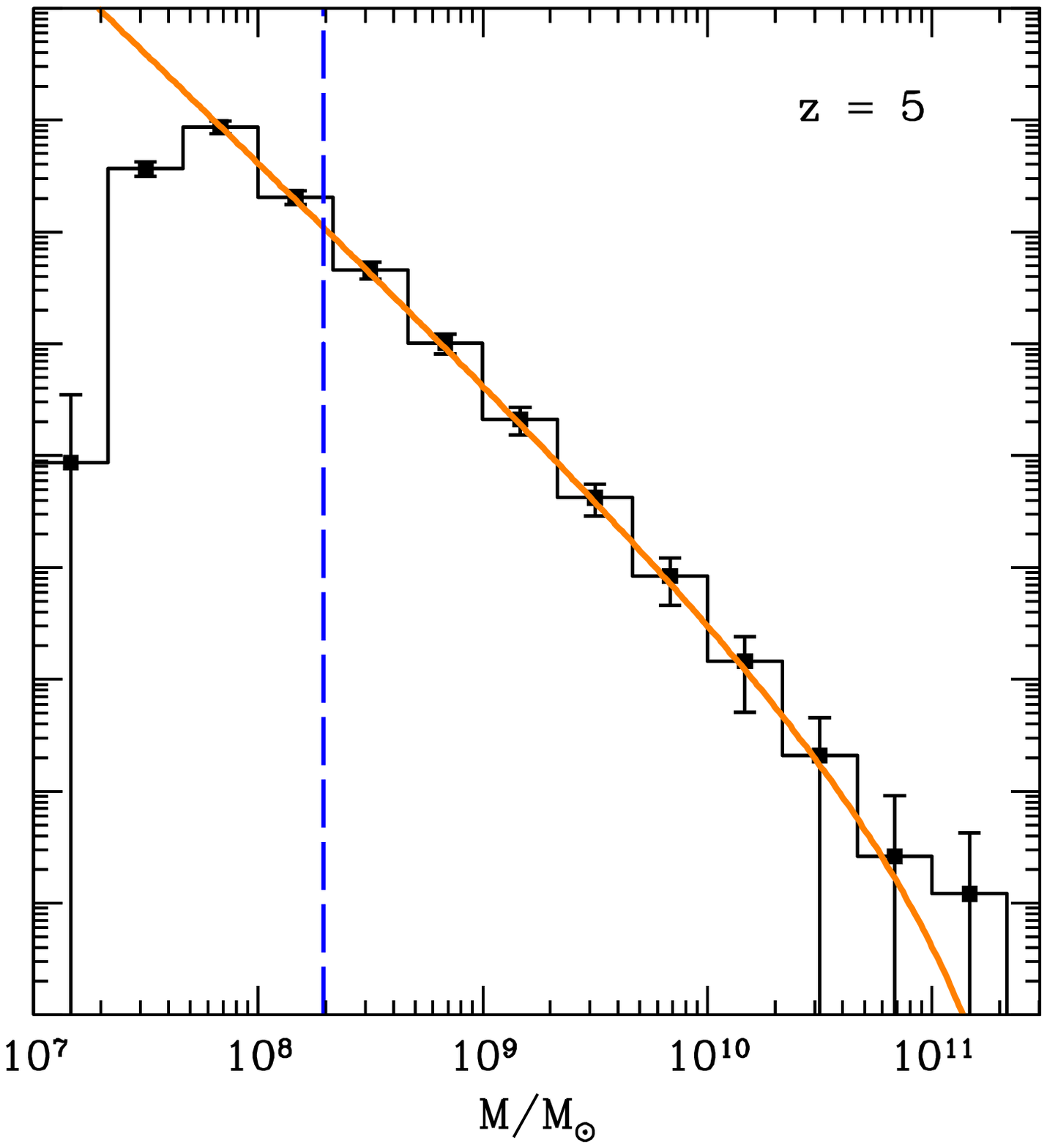}
\hspace{-1.87cm}
\includegraphics [width=5.8cm]{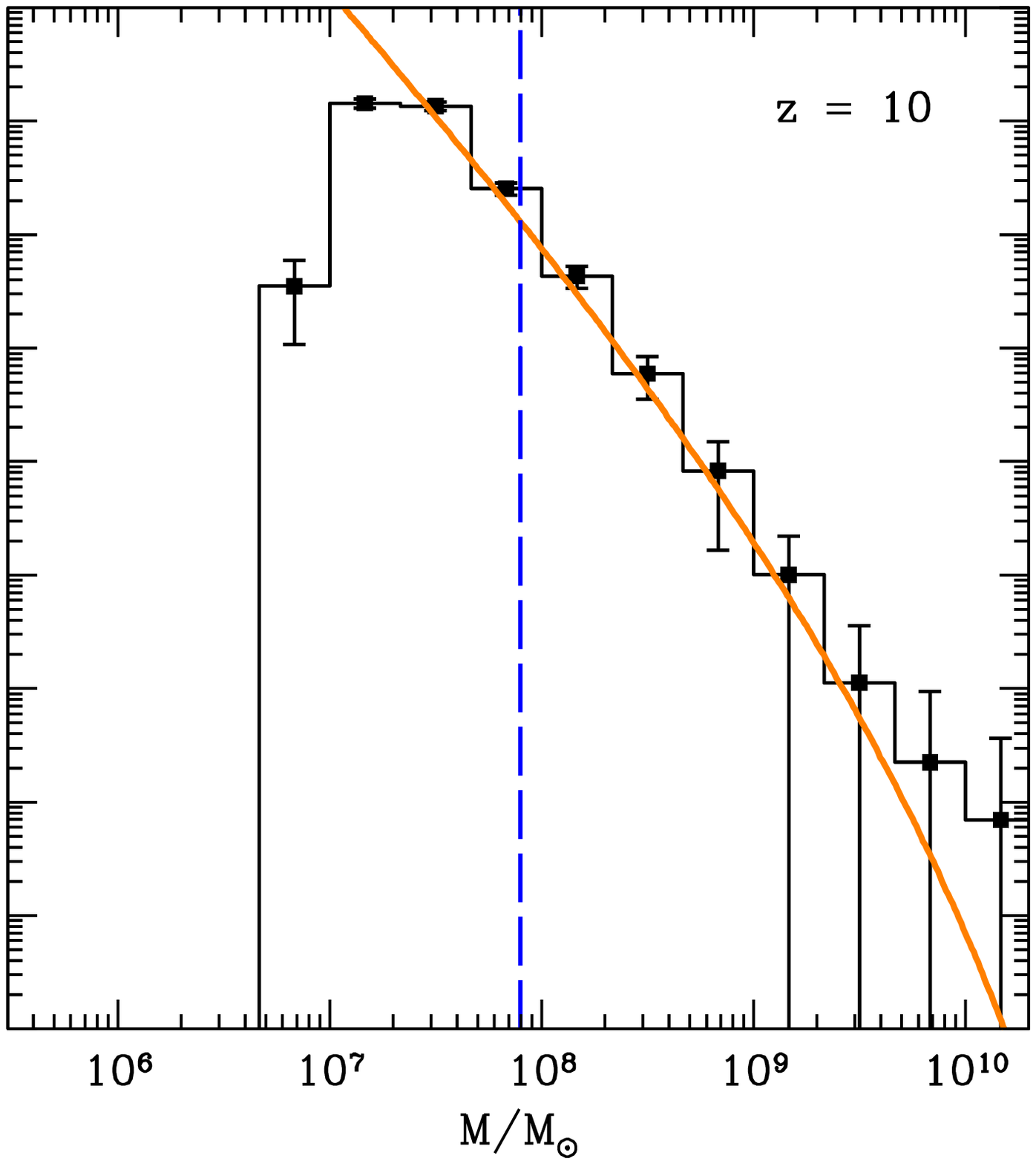}
\hspace{-1.87cm}
\includegraphics [width=5.8cm]{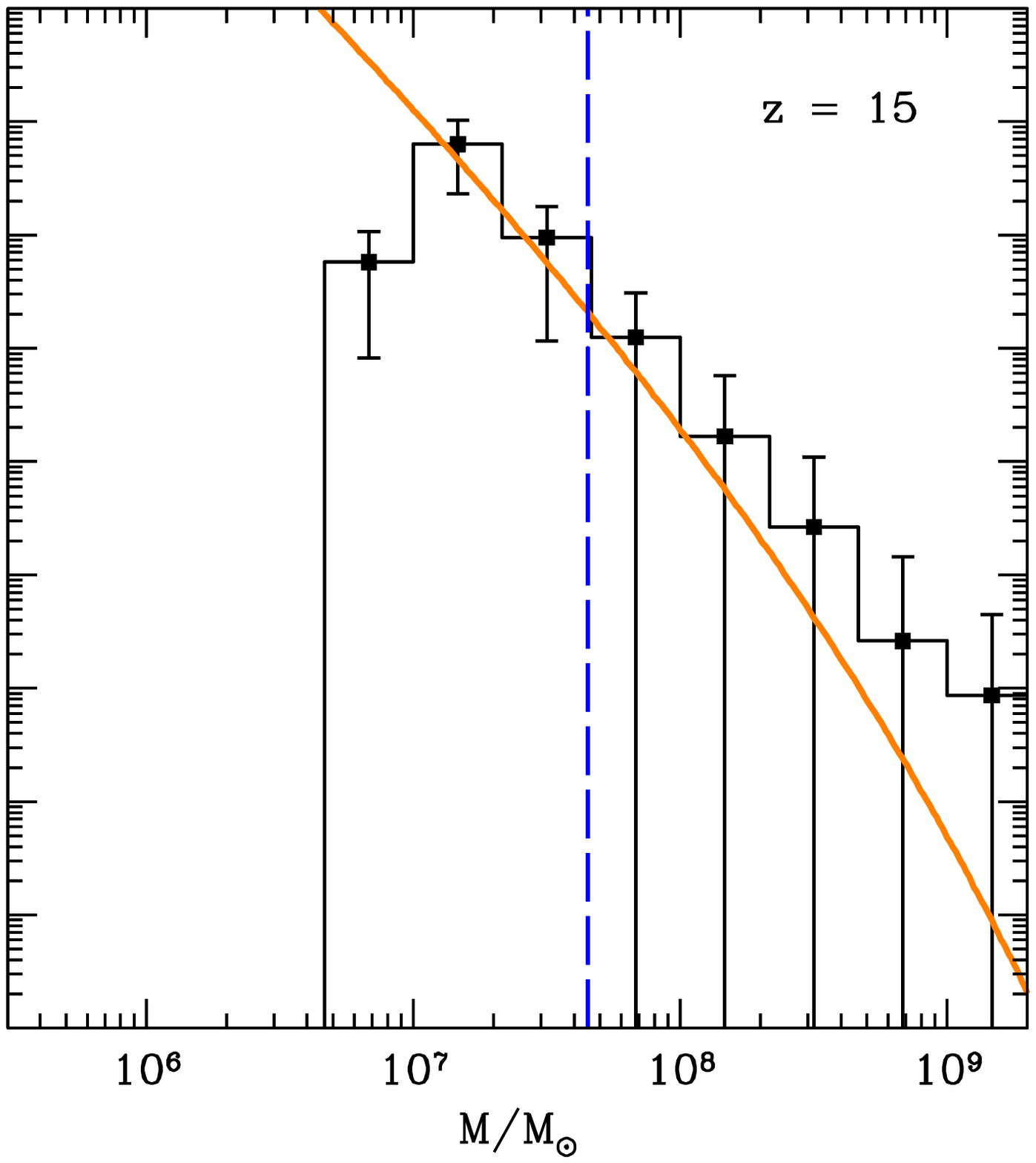}
\caption{Number of halos as function of halo mass at different redshifts (z=1, 5, 10, 15 from left to right) for a MW-like final DM halo 
of $10^{12}$ M$_{\odot}$. The orange solid lines are the analytic predictions from the EPS theory, while the histograms are the simulated mass functions averaged over 50 merger trees. Error bars are given by Poissonian errors. Vertical lines correspond to the adopted minimum mass of star forming halos, $\rm M_{sf}$, at the corresponding redshift.} 
\label{fig:eps} 
\end{figure*} 
%%%%%%%%%%%%%%%%%%%%%%%%%%%%%%%%%%%%%%%%%%%%%%%%%%%%%%%%%

It is clear from the above considerations that stellar archaeology of the most metal-poor stars represents a promising way to explore the
first phases of star formation and metal enrichment in the Universe, constraining the conditions for the formation of the first low-mass stars.
In particular, the low-metallicity tail of the metallicity distribution function (MDF, i.e. the number distribution of stars as function of their [Fe/H]\footnote{$\rm [X/Y] = log_{10}(N_X/N_Y) - log_{10}(N_X/N_Y)_\odot$, for elements X and Y. We adopt solar abundances from Caffau et al. (2011a).}), at $\rm [Fe/H] < -2$ and the observed surface elemental abundances of the most metal-poor stars 
can provide important observational constraints on the nature of Pop III stars and on the formation efficiency of the first Pop II stars (Tumlinson 2006; Karlsson 2006; Salvadori, Schneider \& Ferrara 2007; Salvadori, Ferrara \& Schneider 2008; Salvadori \& Ferrara 2009; Komiya et al. 2009; Salvadori et al. 2010; Komiya 2011).
 A large sample of stars in the MW halo and its satellites has been collected by the HK survey (Beers, Preston \& Schectman 1992), by the Hamburg/ESO survey (HES, Wisotzki et al. 2000, Christlieb 2003), by the Sloan Digital Sky Survey (SDSS, York et al. 2000, Gunn et al. 2006) and by the Sloan Extension for Galactic Understanding and Exploration (SEGUE, Yanny et al. 2009). Many studies have been performed on these data-sets, to assess their completeness (Christlieb et al. 2008, Sch$\rm \ddot{o}$rck et al. 2009), to study the Carbon-enhncement degree (Carollo et al. 2012; Yong et al. 2013b; Norris et al. 2013; Spite et al. 2013) and the chemical abundances of metal-poor stars (see the reviews of Beers \& Christlieb 2005; Frebel \& Norris 2013; Karlsson, Bromm \& 
Bland-Hawthorn 2013 and references therein).
These observations show that the MDF suddenly drops at [Fe/H] $\approx -4$ so that only five stars are observed at [Fe/H] $\lesssim$ -4.75, among which the recently discovered SMSS J031300.36-670839.3 (hereafter SMSS J031300) with $\rm [Fe/H] < -7.1$ (Keller et al. 2014). In general, metal-poor stars in the MW halo have a well defined abundance pattern at $\rm [Fe/H] < - 2.5$ (Cayrel et al. 2004),  that is compatible with theoretical yields of Pop III core-collapse SNe with progenitor masses in the range 
$\rm 10 - 100 \,M_{\odot}$ (Heger \& Woosley 2010; Limongi \& Chieffi 2012). Yet, there is a class of stars, the 
so-called Carbon-enhanced extremely metal-poor stars (CEMP, defined as those with $\rm [C/Fe] >
+1$, Beers \& Christlieb 2005), whose fraction grows with decreasing [Fe/H], being (10 - 20)\% at $\rm [Fe/H] < - 2$ 
(Yong et al. 2013b) and reaching 80\% at $\rm [Fe/H] < - 4.75$. The origin of CEMP stars is 
still debated and many alternative explanations have been proposed 
(see Norris et al. 2013 for a recent critical assessment of the possible formation pathways).

In this paper, we describe an improved version of the hierarchical semi-analytical code  
GA\textsc{laxy} ME\textsc{rger} T\textsc{ree and} E\textsc{volution} (\textsc{gamete}) originally
developed by Salvadori et al. (2007). Our main goal is to investigate whether current observations of
the MDF and of the relative contribution of CEMP stars at different $\rm [Fe/H]$ can provide constraints
on the IMF of Pop III stars and on the physics governing the Pop III/II transition. To this aim, following
Valiante et al. (2011), we have developed a chemical evolution model with dust in a 2-phase interstellar 
medium (ISM); in this model, grains are destroyed by SN shocks in the diffuse, hot phase, but are shielded 
from destruction in the dense, cold phase, where they can grow in mass by accreting gas-phase metals. 
Using this model, we can predict the stellar, gas, metal and dust content of all the MW progenitors along the
simulated merger trees from $z = 20$ down to $z = 0$. With the aim of constraining the model free parameters,
we compare the theoretical predictions with observations of the global properties of the MW, such as the 
gas mass, the stellar content, the metallicity, and the dust-to-gas ratio. 
In addition, we show that the properties of the MW progenitors predicted by the
model appear to be in good agreement with the observed scaling relations between the dust/metal and gas/stellar components
inferred from samples of galaxies at $0 < z < 6$. 

The paper is organized as follows: in Section \ref{sec:model} we give a short description of the basic properties of the \textsc{gamete} 
code, with particular attention to the new features implemented in this work. In Section \ref{sec:calibration} we explain how we calibrate the model free parameters to reproduce the observed global properties of the MW. In Section \ref{sec:dusty} we compare the 
predicted properties of MW progenitors with observational data collected from various galaxy samples. In Section \ref{sec:arch} we apply the model to Galactic Archaeology, exploring the dependence of the simulated MDF to the assumed Pop III IMF (\ref{sec:imf}) 
and Pop III/II transition criteria (\ref{sec:transition}); we make specific predictions for the metallicity distribution of CEMP stars
(\ref{sec:cemp}), comparing the model with observations. Finally, in Sections \ref{sec:disc} and\ref{sec:conc} we discuss our results and present our main conclusions.

\section{Description of the model}
\label{sec:model}

The \textsc{gamete} code allows to reproduce the observed global properties  of the MW, such as the mass of gas and stars, metallicity and dust-to-gas ratio, predicting their evolution along cosmic time. Enrichment processes are assumed to be regulated by stars, both Asymptotic Giant Branch (AGB) stars and SNe, which eject metals and dust into the ISM according to their stellar lifetime (i.e. stars with different masses and metallicities evolve on different timescales); at each time, the star formation rate (SFR) is taken to be 
proportional to the available mass of gas in the ISM, $\rm SFR(t)  = \epsilon_\ast \ M_{ISM}(t)/ t_{dyn}(t)$, where $\rm t_{dyn}(t)$ is the dynamical timescale of the host Dark Matter (DM) halo, and $\epsilon_\ast$ is the star formation 
efficiency, a free parameter of the model.

We follow the formation and evolution of a MW-like galaxy in the framework of the Lamdba Cold Dark Matter ($\Lambda$CDM) model.
Cosmological parameters are taken from the recent results of the Planck Collaboration XVI (2014): h$_0$ = 0.67; $\rm \Omega_b\,h^2$ = 0.022; $\rm \Omega_m$ = 0.32; $\rm \Omega_\Lambda$ = 0.68; $\rm \sigma_8$ = 0.83; $\rm n_s$ = 0.96.
Galaxies are assembled via galaxy-galaxy mergers and/or mass accretion. The hierarchical merger histories (merger trees) of the DM halos  are reconstructed using a binary Monte Carlo algorithm based on the Extended-Press and Schechter (EPS, Press \& Schechter 1974) theory. Here, we briefly summarize the main features of this method and refer the reader to Salvadori et al. (2007, 2008) for 
further details. The code traces the merger tree backward in time, starting from a MW-like DM halo of $\rm M_{MW} = 10^{12} M_{\odot}$ at redshift z = 0 (Binney \& Merrifield 1998; Klypin et al. 2002) up to z $\sim$ 20. At each redshift, DM halos can either fragment in two smaller halos, called \textit{progenitors} and/or loose mass. DM progenitors are identified as halos with masses larger than the so-called resolution mass, $\rm M_{res}$. It is customary in merger-tree models to define the mass below
$\rm M_{res}$ that is not part of collapsed objects as the external or Galactic Medium (GM). This is the environment out of which halos virialize. The free parameters which describe the merger tree model are the redshift interval (dz) and the resolution mass, which are chosen in order to ensure the binarity of the code and the agreement between the resulting mass function of progenitor halos and the theoretically predicted EPS distribution over the relevant redshift range. 

We also assume that star formation in mini-halos is suppressed by strong radiative feedback effects (Ciardi \& Ferrara 2005). Hence, star formation occurs in halos above a minimum mass, $\rm M_{sf} $, that corresponds to the minimum
mass of Ly-$\alpha$
cooling halos, hence to halos with virial temperature $\rm T_{vir} = 2 \times 10^{4}$ K,  
$\rm M_{sf}(z) = M_{vir}(2 \times 10^4 K, z)$ (Barkana \& Loeb 2001).
%\begin{equation}
%\rm
%M_{vir}(z)= 10^8 h^{-1} \Big( \frac{T_{vir}}{2\times 10^4 K} \Big)^{3/2} \Big( \frac{0.6}{\mu} \Big)^{3/2} \Big[  \frac{\Omega_m^z}
%{\Omega_m} \frac{18\pi^2}{\Delta_c} \Big]^{1/2} \Big( \frac{10}{1+z} \Big)^{3/2} M_{\odot}.
%\end{equation}
To properly sample the low-mass  end of the simulated progenitor mass function, we adopt a resolution mass of $\rm M_{res}(z) = M_{sf}(z)/10$.
Fig. \ref{fig:eps} shows the comparison between the theoretical EPS DM halo mass function and the simulated one averaged over 50 merger trees. Different panels correspond to different redshifts and in each panel the vertical line represents the $\rm M_{sf}(z)$ computed at the corresponding redshift. The comparison shows that there is a good agreement with the EPS (within the error bars), particularly for $\rm M \gtrsim M_{sf}(z)$, which is the mass range we are interested in. 

An increase in the minimum mass of star forming halos is expected during cosmic reionization, due to the increased Jeans mass in ionized regions, which causes the quenching of the gas infall in halos below a given circular velocity $\rm v_c(z)$ or DM halo mass (Gnedin 2000; Hoeft et al. 2006; Okamoto, Gao \& Theuns 2008).
While the details of the evolution of $\rm v_c(z)$ depend on the reionization redshift, amplitude of the ionizing
background and redshift (Noh \& McQuinn 2014), here we simply assume that  star
formation is quenched in DM halos with circular velocities smaller than $\rm
v_c = 30\, km\, s^{-1}$ when $\rm z < z_{reion} = 6$ (Salvadori \& Ferrara 2009).

%%%%%%%%%%%%%%%%%%TABLE OBS PROP %%%%%%%%%%%%%%%%%%%%%
\begin{table*}
\begin{center}
  \caption{Observational properties of the MW that we have used to calibrate the free parameters of the model. The existing total mass in metals and dust are derived using values in this Table.
Values are taken from: (1) Stahler \& Palla (2004); (2) Dehnen \& Binney (1998), Brown et al. (2005); (3, 4) Planck Collaboration XXV (2011); (5) Salvadori et al. (2007); (6)  Ganguly et al. (2005); (7) Jenkins (2009); 
(8) Chomiuk \& Povich (2011).}
  \label{tab:masses}
  \begin{tabular}{|cccccccc|}
    \hline
 $\rm M_{ISM}^{(1)}$  &  $\rm M_\ast^{(2)}$  &    $\rm M_d^{diff,(3)}$     &   $\rm M_d^{MC,(4)}$   & $\rm Z_{ISM}^{(5)}$ & $\rm Z_{GM}^{(6)}$ & $\rm {DtM}_{diff}^{(7)}$ & $\rm SFR^{(8)}$ \\
 \hline
(8.0 $\pm$ 2.1) $\times$ $10^{9} M_\odot$ & (6.3 $\pm$ 1.7) $\times$ $10^{10} M_\odot$  &   (8.07 $\pm$ 4.56) $\times$ $10^{7} M_\odot$ &  (6.00 $\pm$ 3.78) $\times$ $10^{7} M_\odot$ &  Z$_\odot$ & 0.25 Z$_\odot$ &  0.57 $\pm$ 0.57 & 1 - 10 $\rm M_\odot/yr$ \\
    \hline
  \end{tabular}
\end{center}
\end{table*}
%%%%%%%%%%%%%%%%%%%%%%%%%%%%%%%%%%%%%%%

\subsection{Two-phase model for the ISM}
\label{sec:new}

We have further implemented \textsc{gamete} to describe the evolution of the gas in two separate phases of the ISM: a \textit{diffuse} 
component (warm/hot low-density gas), where dust can be destroyed by SN shocks, and a \textit{dense} or \textit{molecular cloud} (MC) component (cold and dense gas), where star formation occurs and where dust grains can grow in mass by accreting gas-phase metals
being shielded from destructive processes. While grain reprocessing in the ISM is still subject to many uncertainties, 
observations indicate that dust destruction takes place in regions of the ISM
shocked to velocities of the order of 50 - 200 km s$^{-1}$ (Welty et al. 2002; Podio et al. 2006; 
Slavin 2009). In our model, part of the newly formed dust in SNe is destroyed in situ (hence in the cold dense medium)
by the SN reverse shock (Bianchi \& Schneider 2007). Note also that in the dense phase of the ISM, 
the grains can grow to larger sizes and hence are generally more resistant to sputtering.

 For each quantity (gas, metals, dust), the total mass is simply given by the sum of the masses in the two components. The gas present in the MW environment or GM, where halos are embedded, is assumed to be only in the diffuse phase; 
when a DM halo first virializes, the ISM is initially accreted in the diffuse phase, with a gas infall rate $\rm \dot{M}_{inf}$ (Salvadori et al. 2008).
For star forming systems, with $\rm M_{halo} \ge M_{sf}$, the following processes are considered:
\begin{enumerate}
\item the diffuse gas, $\rm M_{ISM}^{diff}$, condenses into the dense molecular phase, $\rm M_{ISM}^{MC}$, at a rate $\rm \dot{M}_{\rm cond}$;
\item stars ($\rm M_\ast$) form out of dense gas with a star formation rate SFR;
\item stars evolve and return gas, metals and dust in the diffuse phase;
\item dust in molecular clouds accretes gas-phase metals, while dust in the diffuse phase is partly destroyed by interstellar shocks due to SN explosions;
\item mechanical feedback due to SN explosions drives gas outflows, ejecting the diffuse gas into the GM; 
%$\rm M_{ej}^{diff}\ and\ M_{ej}^{MC}$;
\item part of the dense phase returns to the diffuse phase due to dispersion of molecular clouds.
\end{enumerate}
As a result, the time evolution of the gas is described by the following set of differential equations:
\begin{equation}
\label{eq:ismdiff}
  \rm
  \dot{M}_{ISM}^{diff}(t) = \dot{M}_{inf}(t) - \dot{M}_{cond}(t)+\dot{R}(t) -\dot{M}_{ej}^{diff}(t),
\end{equation}
\noindent
and
\begin{equation}
\label{eq:ismmc}
\rm  
 \dot{M}_{ISM}^{MC}(t) = \dot{M}_{cond}(t) -SFR(t) - \dot{M}_{ej}^{MC}(t).
\end{equation}
\noindent
In the above expressions, $\rm \dot{R}(t)$ is the gas returned by stars to the ISM (Salvadori et al. 2008), and $\rm \dot{M}_{ej}^{diff}(t)$ and $\rm \dot{M}_{ej}^{MC}(t)$ represent the diffuse and dense gas ejection rate.
The total mass of gas ejected per unit time $\rm \dot{M}_{ej}$ due to SN-driven mechanical feedback is proportional 
to the SN explosion rate and the wind energy is regulated by a free parameter $\rm \epsilon_w$, controlling the conversion efficiency of SN explosion energy into kinetic energy (see Eq. 9 of Salvadori et al. 2008). 
Since the effect of SN-driven on different ISM phases is still poorly understood (Efstathiou 2000, Powell et al. 2011), we studied different mass-loading prescriptions, finding that the higher is the MC ejection rate, the higher must be the condensation rate to reproduce the global properties of the MW.
Hence we decide to follow Fu et al. (2013) and consider that the energy injected into the medium from SNe leads to the ejection only of the hot diffuse phase through mechanical feedback (i.e. $\rm \dot{M}_{ej}^{MC}(t) = 0$).

The physical processes leading to the formation of  molecular clouds are still not fully understood (see Dobbs et al. 2013 for a recent review). Here we adopt a simple empirical description, tailored to reproduce the observed mass of gas in the molecular and atomic components of the MW. 

We assume the condensation rate to be proportional to the star formation rate, hence 
$\rm \dot{M}_{cond}(t) = \alpha_{MC}\, SFR(t)$, where $\rm \alpha_{MC}$ is a free parameter of the model.
% where $\rm \alpha_{MC} > 1$ to ensure the formation of dense gas and stars.
The above condensation rate is the net result of the mass exchange between the two 
ISM phases, which is regulated by the condensation of the
diffuse gas into molecular clouds, 
\begin{equation}
\rm 
 M_{\rm ISM}^{\rm diff}(t)/t_{form} = \frac{1}{1-\textit{f}} \, \, \alpha_{MC} \,SFR(t), 
\end{equation}
\noindent
and the dispersal of molecular gas into the diffuse phase,
\begin{equation}
\rm 
 M_{\rm ISM}^{\rm MC}(t)/t_{des} = \frac{\textit{f}}{1-\textit{f}} \, \, \alpha_{MC} \, SFR(t),
 \end{equation}
\noindent
where 0 $< \textit{f} <$ 1. Hence, the two-phase structure of the ISM is regulated by the
free parameters \textit{f} and $\rm \alpha_{MC}$. When $f \rightarrow$ 0, the condensation process 
is very efficient and the gas content in the ISM is dominated by the MC phase. 
Conversely, when $f \rightarrow$ 1 the condensation and dispersal rates are in equilibrium. 
As a result, the newly condensed phase is rapidly dispersed in the ISM, preventing the formation of a stable dense phase. 
As we will show in Section \ref{sec:calibration}, the values of the two parameters, $\rm \alpha_{MC}$ and $\textit{f}$, have been selected 
to reproduce the observed molecular and atomic gas masses in the MW.

This two-phase model has been recently applied to high redshift quasars (Valiante et al., 2014), where accretion onto the central black hole and Active Galactic Nuclei (AGN) feedback are also included.

%%%%%%%%%%%%%%%%%%%%%%%%%%%%%%%%%%%%%%%%%%%%%%%%%%%%%%%%%
%%% Figure integrated yields
\begin{figure*}
\vspace{-4.cm}
\hspace{-1.5cm}
\includegraphics [width=19.0cm]{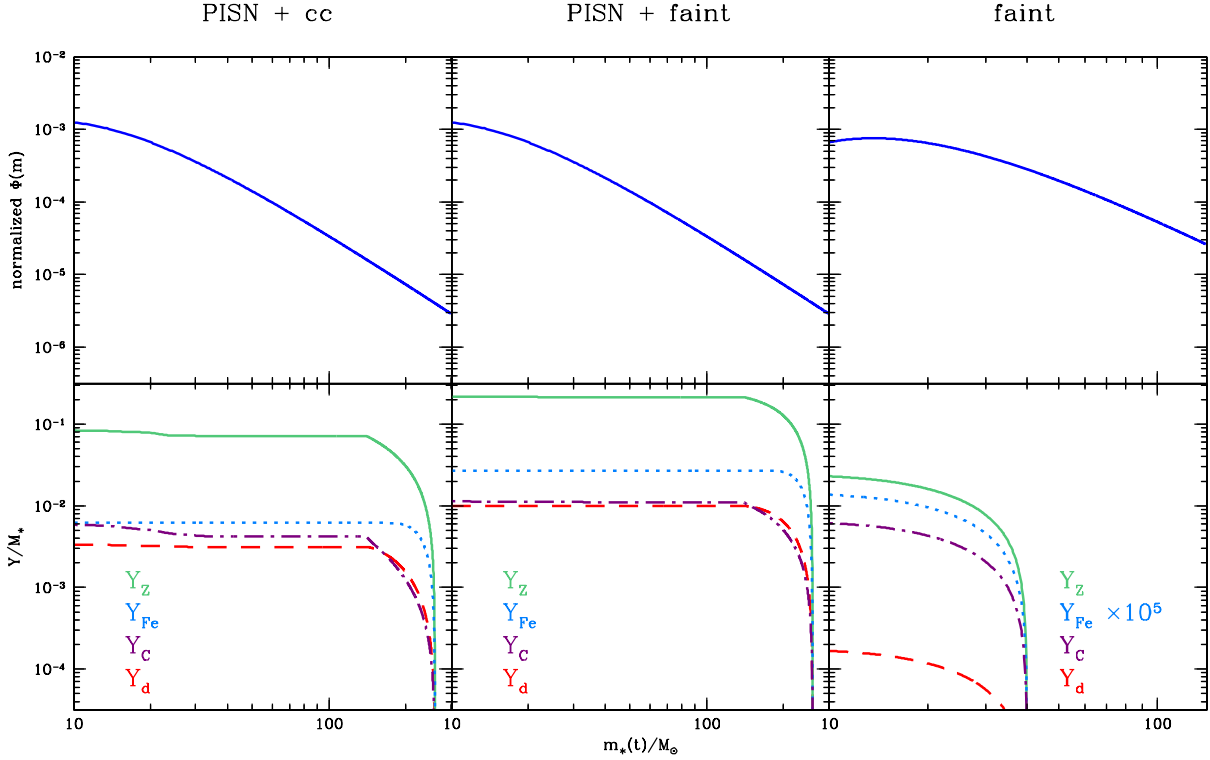}
\vspace{-5.cm}
\caption{Description of the Pop III models explored. In the upper panels we show the normalized Pop III IMF that we consider and in the bottom panel
the corresponding IMF-integrated yields. These are normalized to the stellar mass formed in a single burst and plotted as a function of the minimum stellar mass (see Eqs. \ref{eq:myields} and \ref{eq:dyields}). We show the separate evolution of the total metal (green, solid) and dust (red, dashed) yields and the
evolution of the iron (azure, dotted) and carbon (purple, dot-dashed) yields. Left panels: stars form in the mass range [10-300] $\rm M_\odot$ assuming cc SNe yields for masses m $<$ 40 M$_\odot$. Middle panels: stars form in the same mass range but faint SNe yields are considered if m $<$ 40 M$_\odot$. Right panels: stars form in the mass range [10-140] M$_\odot$, excluding the PISN mass range and considering faint SNe yields. Note that in this last case the iron yield is multiplied by a factor $10^5$.}
\label{fig:yieldsIII}
\end{figure*}
%%%%%%%%%%%%%%%%%%%%%%%%%%%%%%%%%%%%%%%%%%%%%%%%%%%%%%%%%

\subsection{Chemical evolution}
\label{sec:chem}

The enrichment in heavy elements (including both gas-phase metals and metals locked in dust grains) is described in the two-phase ISM by means of the following equations:
\begin{eqnarray}
  \rm
  \dot{M}_{Z}^{\rm diff}(t) & = &\rm Z_{vir}(t) \, \dot{M}_{inf}(t) \rm -Z_{diff}(t) \,M_{\rm ISM}^{\rm diff}(t)/t_{form} 
   \nonumber \\
  & & \rm + Z_{MC}(t) \,M_{\rm ISM}^{\rm MC}(t)/t_{des}+\dot{Y}_Z(t) -Z_{diff}(t) \, \dot{M}_{ej}^{\rm diff}(t),
\end{eqnarray}
\noindent
and 
\begin{eqnarray}
  \rm 
  \dot{M}_{Z}^{\rm MC}(t) & = & \rm Z_{diff}(t) \,M_{\rm ISM}^{\rm diff}(t)/t_{form} - Z_{MC}(t) \,M_{\rm ISM}^{\rm MC}(t)/t_{des} \nonumber \\ 
   & & \rm  - Z_{MC}(t)\, \rmn{SFR}(t) -Z_{MC}(t)\, \dot{M}_{ej}^{\rm MC}(t),
\end{eqnarray}
where $\rm Z_{diff} = M_Z^{\rm diff}(t)/M_{\rm ISM}^{\rm diff}(t)$ and $\rm Z_{MC} = M_Z^{\rm MC}(t)/M_{\rm ISM}^{\rm MC}(t)$ are the total (gas-phase + dust) metallicities of the diffuse and dense phases, $\rm Z_{vir}$ is the metallicity of the gas accreted from the GM and $\rm \dot{Y}_Z(t)$ is the rate at which heavy elements produced by stars are returned to the ISM, given by,
\begin{equation}
\label{eq:myields}
\rm \dot{Y}_Z(t) = \rm \int_{m_\ast(t)}^{m_{up}} m_Z(m,Z)\ \Phi(m)\ SFR(t - \tau_m)\ dm,
\end{equation}
\noindent where the lower limit of integration refers to the mass of a star with a lifetime $\rm \tau_m$ = t,  $\rm m_Z(m,Z)$ is the total mass of metals (pre-existing and newly synthesized) produced by a star of initial mass m and metallicity Z, $\Phi$(m) is the stellar IMF and $\rm m_{up}$ is the upper mass limit of the IMF.

The chemical network has been extended to implement the evolution of dust (Valiante et al. 2009, 2011). In the two phase ISM, this is described as,
\begin{eqnarray}
\rm 
  \dot{M}_{d}^{\rm diff}(t) & = & \rm \mathcal{D}_{vir}(t)\dot{M}_{inf}(t) -\mathcal{D}_{diff}(t)\,
  M_{\rm ISM}^{\rm diff}(t)/t_{form} - \rm M_{d}^{diff}(t)/\tau_d   \nonumber \\
  & & \rm+\mathcal{D}_{MC}(t) \, M_{\rm ISM}^{\rm MC}(t)/t_{des}+\dot{Y}_d(t)-\mathcal{D}_{diff}(t)\dot{M}_{ej}^{\rm diff}(t),
\end{eqnarray}
\noindent
and 
\begin{eqnarray}
\label{eq:mdmc}
 \rm 
  \dot{M}_{d}^{\rm MC}(t) & = & \rm \mathcal{D}_{diff}(t)\,M_{\rm ISM}^{\rm diff}(t)/t_{form}  -\mathcal{D}_{MC}(t) \, M_{\rm ISM}^{\rm MC}(t)/t_{des} 
    \nonumber \\ 
  & & \rm -\mathcal{D}_{MC}(t)\rmn{SFR}(t) + M_{d}^{MC}/\tau_{acc}-\mathcal{D}_{MC}(t)\dot{M}_{ej}^{\rm MC}(t),
\end{eqnarray}
\noindent
where $\rm \mathcal{D}_{diff}(t)=M_{d}^{\rm diff}(t)/M_{\rm ISM}^{\rm diff}(t)$ and $\rm \mathcal{D}_{MC}(t)=M_{d}^{\rm MC}(t)/M_{\rm ISM}^{\rm MC}(t)$ are the dust-to-gas (mass) ratios in the two phases, $\rm \mathcal{D}_{vir}$ is the dust-to-gas ratio of the infalling gas and $\rm \dot{Y}_d$ is the rate at which dust produced by stars is injected into the ISM by AGB stars and SN explosions (Valiante et al. 2009),
\begin{equation}
\label{eq:dyields}
\rm \dot{Y}_d(t) = \rm \int_{m_\ast(t)}^{m_{up}} m_d(m,Z)\ \Phi(m)\ SFR(t - \tau_m)\ dm
\end{equation}
\noindent where $\rm m_d(m,Z)$ is the mass of dust produced by a star with mass m and metallicity Z, and all other quantities are the same as in Eq. \ref{eq:myields}.\\
For any given element/grain species $i$, the corresponding yield is computed as in Eqs. \ref{eq:myields} and \ref{eq:dyields} substituting $\rm m_Z$ and $\rm m_d$ with $\rm m_i(m,Z)$. A presentation of the mass- and metallicity-dependent dust and metal yields will be given in Sections \ref{sec:popIIIimf} and \ref{sec:popIIimf}.

%%%%%%%%%%%%%%%%%%%%%%%%%%%%%%%%%%%%%%%%%%%%%%%%%%%%%%%%%
%%% Figure integrated yields
\begin{figure}
\vspace{-4.0cm}
\hspace{-1.5cm}
\includegraphics [width=19.0cm]{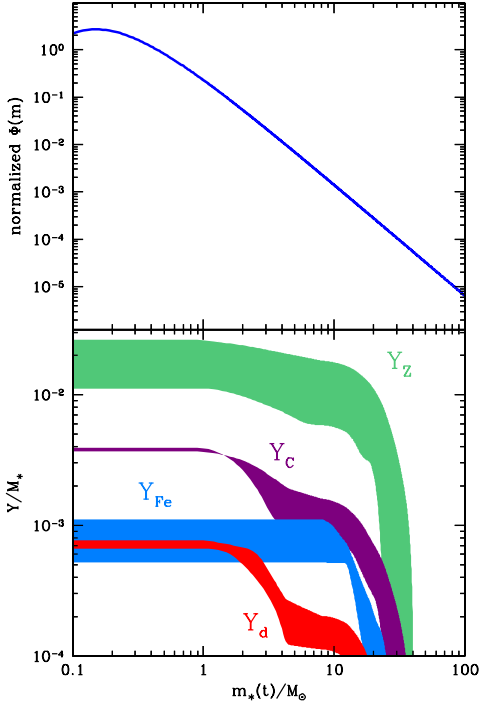}
\vspace{-5.5cm}
\caption{Same as Fig.\ref{fig:yieldsIII} but for Pop II stars. Shaded areas account for metallicity-dependent yields, with $\rm 10^{-4} Z_\odot < Z < Z_\odot$ (see text).}
\label{fig:yieldsII}
\end{figure}
%%%%%%%%%%%%%%%%%%%%%%%%%%%%%%%%%%%%%%%%%%%%%%%%%%%%%%%%%

Grain growth through accretion of gas-phase metals can occur only in the dense phase of the ISM (see Eq. \ref{eq:mdmc}); following Asano et al. (2013), we model the dust accretion timescale $\rm \tau_{acc}$ as:
\begin{equation}
\label{eq:tauacc}
\rm
\tau_{acc} \approx 20\, Myr\times \Big( \frac{\bar{a}}{0.1\, \mu m} \Big)\, \Big( \frac{n}{100\, cm^{-3}} \Big)^{-1}\, \Big( \frac{T}{50\, K} \Big)^{-3/2}\, \Big( \frac{Z}{Z_\odot} \Big)^{-1}
\end{equation}
\noindent
where $\rm \bar{a}$ is the typical size of dust grains, $\rm n$ is the number density of the gas in the MC phase, T and Z are the temperature and the gas-phase metallicity of the cloud, respectively. Here
we assume that in the MC phase $\rm n = 10^3 cm^{-3}$ and $\rm T = 50 K$, so that the accretion timescale  from a gas at solar metallicity is 2 Myr (Asano et al. 2013).

The destruction timescale, $\rm \tau_d$, is the lifetime of dust grains destroyed by thermal sputtering in high-velocity (v $\rm > 150\, km\, s^{-1}$) SN shocks. We assume that dust grains can be destroyed only in the hot diffuse phase of the ISM and that,
\begin{equation}
\rm
\tau_d = \frac{ M_{ISM}^{diff} }{ \epsilon_d \ M_{swept} \ R'_{SN} },
\end{equation}
\noindent
where $\epsilon_d$ is the destruction efficiency, $\rm M_{swept}$ is the effective ISM mass that is completely cleared of dust by a single SN shock and $\rm R'_{SN} = f_{SN}R_{SN}$ is the effective SN rate for grains destruction, accounting for the possibility that not all the SNe which interact with the ISM are equally efficient at destroying dust (Dwek \& Scalo 1980; McKee 1989; Tielens 2005).
In what follows, we assume that stars with masses in the range $\rm [140-260]\, M_\odot$ explode as Pair-Instability SuperNovae (PISN, Heger \& Woosley 2002). These are very efficient at destroying dust (Nozawa et al. 2006), hence $\rm f_{PISN} = 1$. Conversely, stars with masses in the range $\rm [8-40]\, M_\odot$ explode as core-collapse SNe (either faint or ordinary SNe, see section \ref{sec:popIIIimf}) and are less efficient at destroying dust, $\rm f_{SN} = 0.15$. 
The mass $\rm M_{swept}$ can be computed using the Sedov-Taylor solution for an expanding SN in a homogeneous medium 
with shock velocity $\rm v_{sh}$ (McKee 1989),
\begin{equation}
\rm
M_{swept} = 6800\, M_\odot \frac{\langle E_{SN} \rangle /10^{51}erg}{(v_{sh}/100\, km\, s^{-1})^2},
\end{equation}
\noindent
where $\rm v_{sh} \sim 200\, km\, s^{-1}$ is the (minimum) non-radiative shock velocity and $\rm \langle E_{SN} \rangle$ is the average supernova explosion energy, assumed to be 2.7 $\times 10^{52}$ erg for PISNe and 1.2 $\times 10^{51}$ erg for core collapse SNe. The adopted dust destruction efficiencies are taken from Nozawa et al. (2006) for PISNe and from Jones et al. (1996) for SNe, $\rm \epsilon_d \sim 0.6$ and $\rm \epsilon_d \sim 0.48$ respectively, and assume a reference ISM density of $\rm n_{ISM} = 1\, cm^{-3}$ . 

Note that the mass of gas-phase metals can be easily inferred as $\rm M_Z - M_d$.

\subsection{Population III stars}
\label{sec:popIIIimf}

We assume Pop III stars to form  according to a Larson-type IMF (Larson 1998), 
\begin{equation}
\label{larson}
\rm
\Phi(m) = \frac{dN}{dm} \propto m^{\alpha-1} exp\Big(-\frac{m_{ch}}{m}\Big)
\end{equation}
with $\rm \alpha = -1.35$. The value of $\rm m_{ch}$ is chosen in order to have an average mass of Pop III stars of $\rm \sim 40\, M_\odot$ (Hosokawa \& Omukai 2009). In the first model that we explore,
Pop III stars are assumed to form with masses in the range $\rm [10 - 300] \, M_\odot$ with a characteristic mass of $\rm m_{ch}$ = 20 M$_\odot$.  The corresponding Pop III IMF is shown in the
upper left panel of Fig.~\ref{fig:yieldsIII}.
We follow the gradual enrichment of the ISM by Pop III stars taking into account their mass-dependent lifetimes (Raiteri et al. 1996, Schaerer 2002), and consider metal and dust yields from core-collapse SNe ($\rm 10\, M_\odot < m_\ast < 40\, M_\odot$) and PISNe ($\rm 140\, M_\odot < m_\ast < 260\, M_\odot$); outside these two mass ranges stars are assumed to directly collapse to black hole without ejecting the products of their nucleosynthesis.
For the purpose of the present study, we follow the evolution of the total mass in metals with the separate contributions of Fe, C and O using the yields of Woosley \& Weaver (1995) for core-collapse SN and of  Heger \& Woosley (2002) for PISN. 
In addition, we also follow the enrichment in dust grains produced in the ejecta of Pop III SNe, using the results by Bianchi \& Schneider (2007) for core-collapse SNe and by Schneider et al. (2004) for PISNe. We take into account the partial destruction of the newly formed grains by a reverse shock of moderate intensity (Bianchi \& Schneider 2007), so that approximately 7\% of the newly condensed dust mass is able to survive and enrich the ISM (Valiante et al. 2009). In the bottom left panel of Fig.~\ref{fig:yieldsIII}, we show the Pop III IMF-integrated yields given by Eqs.~\ref{eq:myields} and \ref{eq:dyields} as a function of the minimum stellar mass, $\rm m_\ast(t)$.
Here all the stars are assumed to form in a burst and the total mass ejected is normalized to the total mass of stars formed. 
In the second Pop III model that we explore, we keep the same Pop III IMF and mass range as described above but we assume that 
stars in the $\rm 10\, M_\odot < m_\ast < 40\, M_\odot$ mass range explode as faint SNe, hence are
characterized by moderate mixing and strong fallback (Umeda \& Nomoto 2003). For these explosions, we adopt metal and dust yields recently computed by Marassi et al. (2014) and show the corresponding IMF and IMF-integrated yields in the central panels of Fig.~\ref{fig:yieldsIII}.
It is clear from the figure that there are only minor differences between faint and ordinary core-collapse SNe, since massive PISNe are the first to explode and dominate the yields. Only C shows a larger variation, but the difference between the two models is smaller than a factor $\sim$ 2.
Finally, in the last model that we explore Pop III stars are assumed to form in the mass range $\rm [10 - 140]\,M_\odot$ with a characteristic mass $\rm m_{ch}$ = 32 M$_\odot$ (top right panel), excluding the PISN mass range and assuming faint SNe at m $<$ 40 M$_\odot$. The resulting IMF-integrated yields are show in the bottom-right panel: for stars with 40 M$_\odot$ $<$ m $<$ 140 M$_\odot$ the ejected material completely fallbacks onto the star. Hence stars contribute to the integrated yields only if m $<$ 40 M$_\odot$. While metals/dust/C-yields are a factor $\sim$ 5/20/50 smaller than faint SNe + PISNe case (middle panel), the main feature of faint SNe is the dramatically lower Fe yield ($\sim$ 5 orders of magnitude smaller). A summary of the three Pop III models that we consider is given in Table~\ref{table:models2}.

%%%%%%%%%%%%%%%%%%%%%%%%%%%%%%%%%%%%%%%%%%%%%%%%%%%%%%%%%
\begin{center}
\begin{table}
 \caption{Model parameters adopted in the present study. The first four entries represent common parameters that have been fixed so as to match the observed global 
properties of the MW. The last three entries show the adopted values for the critical parameters that control the Pop III/II transition.}
\begin{tabular}{|c c c c c c c c|}
\hline
\rm
 $\epsilon_\ast$ & $\epsilon_w$ & $\alpha_{MC}$ & $f$ &  $\rm Z_{cr}$&$\rm \mathcal{D}_{cr}$&$\rm D_{trans,cr}$ & \\
\hline 
 0.8 & 0.0016 & 1.02 & 0.4 & $\rm 10^{-3.8}Z_\odot$&$4.4 \times 10^{-9}$&-3.5 & \\  
\hline 
\end{tabular}
\label{table:models1}
\vspace{-1.1cm}
\end{table}
\end{center}
%%%%%%%%%%%%%%%%%%%%%%%%%%%%%%%%%%%%%%%%%%%%%%%%%%%%%%%%%

%%%%%%%%%%%%%%%%%%%%%%%%%%%%%%%%%%%%%%%%%%%%%%%%%%%%%%%%%
%%% Figure masses and SFR evolution for calibration
\begin{figure*}
\hspace{-0.9cm}
\includegraphics [width=6.4cm]{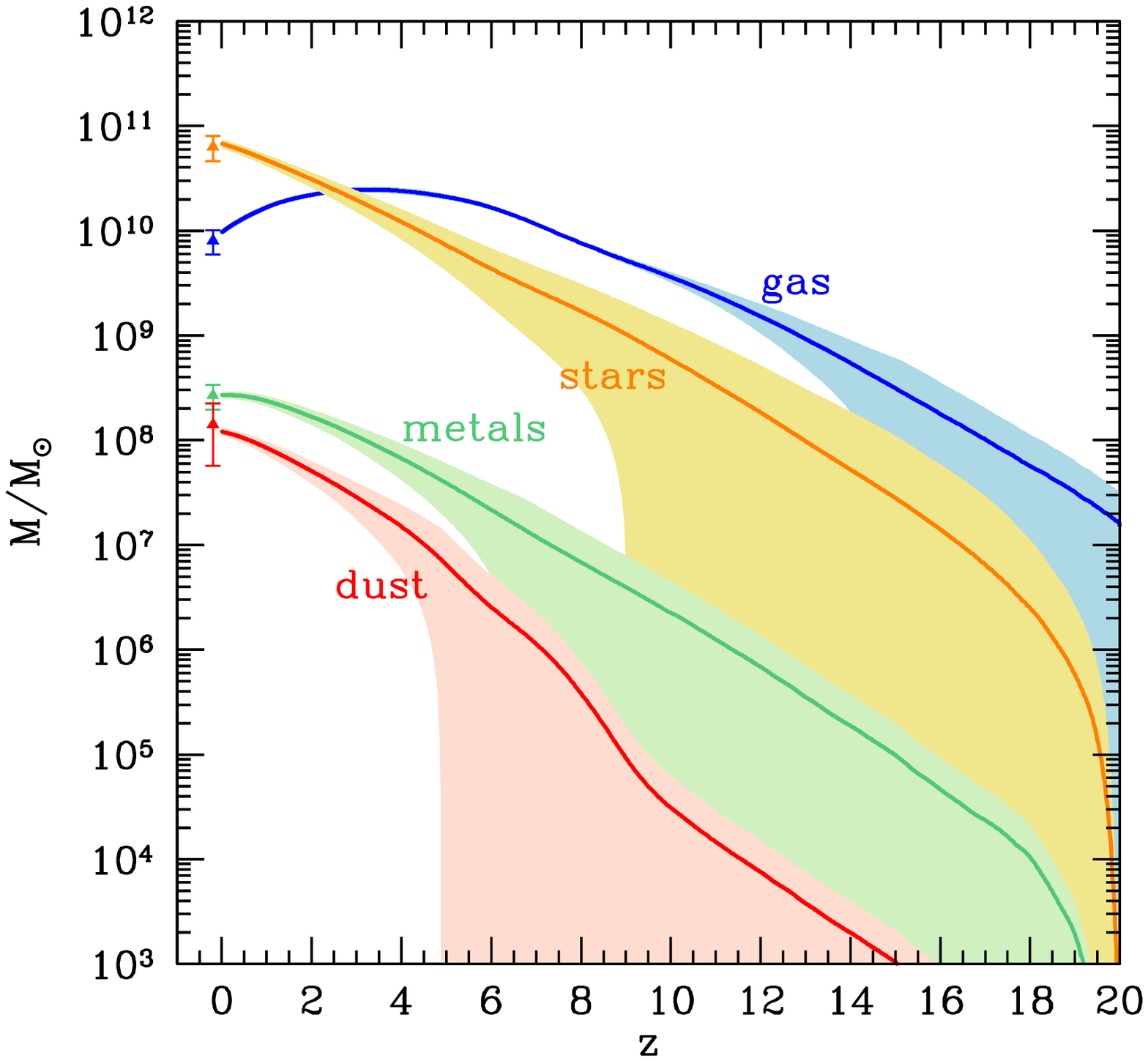}
\hspace{-0.5cm}
\includegraphics [width=6.4cm]{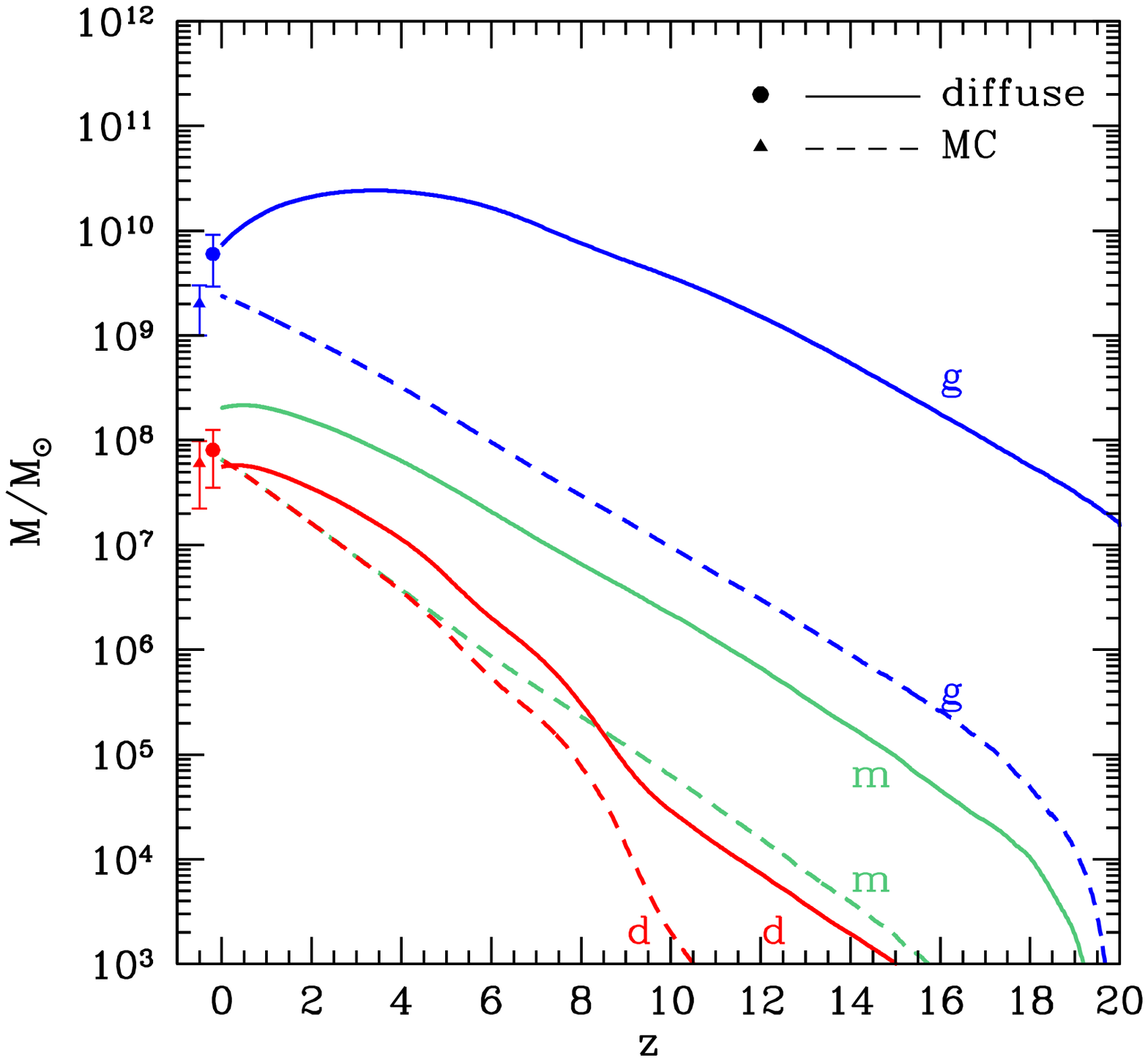}
\hspace{-0.5cm}
\includegraphics [width=6.4cm]{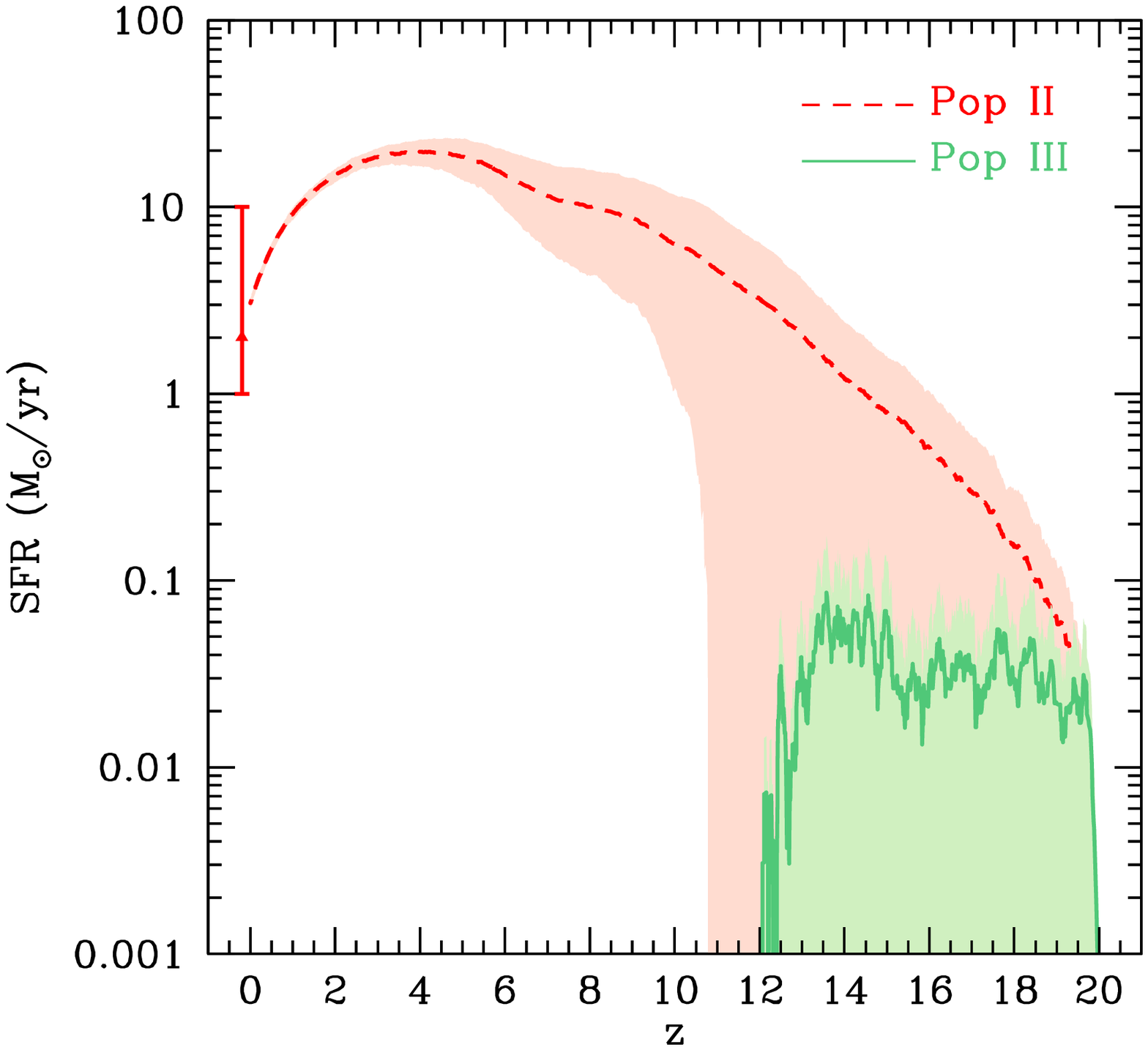}
\vspace{-0.5cm}
\caption{Redshift evolution of the predicted global properties of the MW. Each line represents an average over 50 merger tree realizations with shaded areas representing
$\rm 1-\sigma$ Poissonian errors. {\it Left panel}: gas (blue), star (yellow), metal (green) and dust (red) masses are shown for the fiducial model parameters of Table \ref{table:models1}.  {\it Middle panel}: separate evolution of gas (g), metals (m) and dust (d) in the diffuse (solid lines) and MC (dashed lines) ISM phase. %separate evolution of diffuse (solid lines) and MC (dashed lines) phases for gas, metals and dust. 
Observed quantities at $\rm z = 0$ are also shown (points with errorbars, see Table \ref{tab:masses}). {\it Right panel}: redshift evolution of 
Pop II (dashed red) and Pop III (solid green) star formation rate for model (f) (see text and Table~\ref{table:models2}).}
\label{fig:calib} 
\end{figure*}
%%%%%%%%%%%%%%%%%%%%%%%%%%%%%%%%%%%%%%%%%%%%%%%%%%%%%%%%%

\subsection{The Pop III/II transition}
\label{sec:trans}

As we have discussed in Section~\ref{sec:intro}, one of the goal of the present study is to investigate the potential of current observations of 
metal-poor stars in the MW halo to constrain the physical conditions that enable the formation of the first low-mass Pop II stars. The chemical
evolution model described above allows to follow the evolution of the mass of gas-phase metals, dust grains, as well as the
separate abundances of C, O, and Fe in the 2-phase ISM. Hence, it enables us to investigate the model predictions assuming 
different Pop III/II transition criteria. In particular:
\begin{enumerate}
\item {\it critical metallicity}: low-mass Pop II stars can form when the gas-phase
metallicity of MC is larger than $\rm  Z_{cr} = 10^{-3.8} \, Z_{\odot}$ (Bromm et al. 2001); 
\item {\it critical dust-to-gas ratio}: Pop II stars form when the dust-to-gas mass ratio of MC is larger than $\mathcal{D}_{cr} = 4.4^{+1.9}_{-1.8} \times 10^{-9}$, where the errorbars 
take into account the modulation due to grain properties (Schneider et al. 2012a). 
\item {\it transition discriminant}:  Pop II stars
form if the transition discriminant, $\rm D_{trans} = log_{10}(10^{[C/H]}+0.9 \times 10^{[O/H]})$, where [C/H] and [O/H]
are the total (gas-phase + dust) C and O abundance in the MC, is larger than $\rm D_{trans,cr} = -3.5 \pm 0.2$ (Frebel \& Norris 2013).
\end{enumerate}
Table \ref{table:models2} lists the models that we have explored using different Pop III/II transition criteria.

\subsection{Population II stars}
\label{sec:popIIimf}

Once the criteria for low-mass star formation are met, Pop II stars are assumed to form according to a Larson IMF in the mass range [0.1 - 100] M$_\odot$ with $\rm m_{ch} = 0.35\ M_\odot$ (Larson 1998) and we consider their evolution taking into account the mass- and metallicity-dependent stellar lifetimes (Raiteri et al. 1996).
For stars with $\rm m_\ast < 8\, M_\odot$, we adopt metal yields from Van den Hoeck \& Groenewegen (1997) and dust yields for intermediate-mass stars on the AGB phase of the evolution by Zhukovska et al. (2008). Metal and dust yields for massive stars $\rm (12\, M_\odot < m_\ast < 40\, M_\odot)$ have been taken from Woosley \& Weaver (1995) and from Bianchi \& Schneider (2007), using the proper mass-and metallicity-dependent values. For stars in the mass range $\rm 8\, M_\odot < m_\ast < 12\, M_\odot$, we interpolate between the AGB yields for the largest-mass progenitor and SN yields from the lowest-mass progenitor. Above 40 $\rm M_\odot$, stars are assumed to collapse to black hole without contributing to the enrichment of the ISM. 
The chosen IMF and the resulting Pop II IMF-integrated yields are shown in Fig~\ref{fig:yieldsII} (upper and lower panels, respectively). The shaded areas represent how yields vary for different stellar progenitor metallicities (see also Valiante et al. 2009). Similarly to what has been assumed for Pop III dust yields, we take into account the partial destruction by the
reverse shock of the newly formed dust in SN ejecta; with this choice, the resulting SN dust masses appear to be in good agreement with available observational data of SNe and SN remnants 
(Valiante \& Schneider 2014; Schneider et al. 2014).

%%%%%%%%%%%%%%%%%%%%%%%%%%%%%%%%%%%%%%%%%%%%%%%%%%%%%%%%%
%%% Figure for comparison with Corbelli data (1) and (2)
\begin{figure*}
\hspace{-0.8cm}
\includegraphics [width=6.4cm]{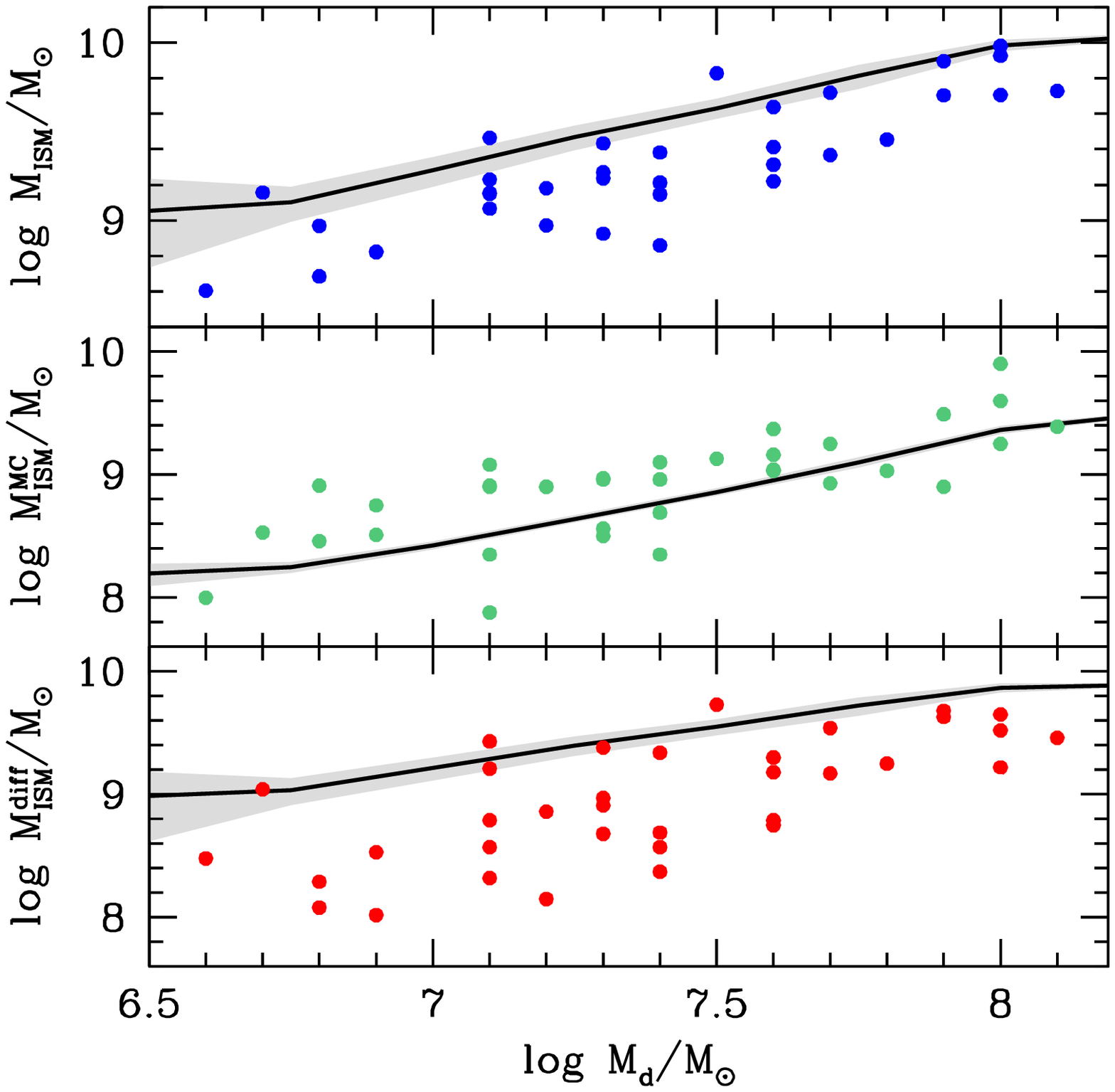}
\hspace{-0.6cm}
\includegraphics [width=6.4cm]{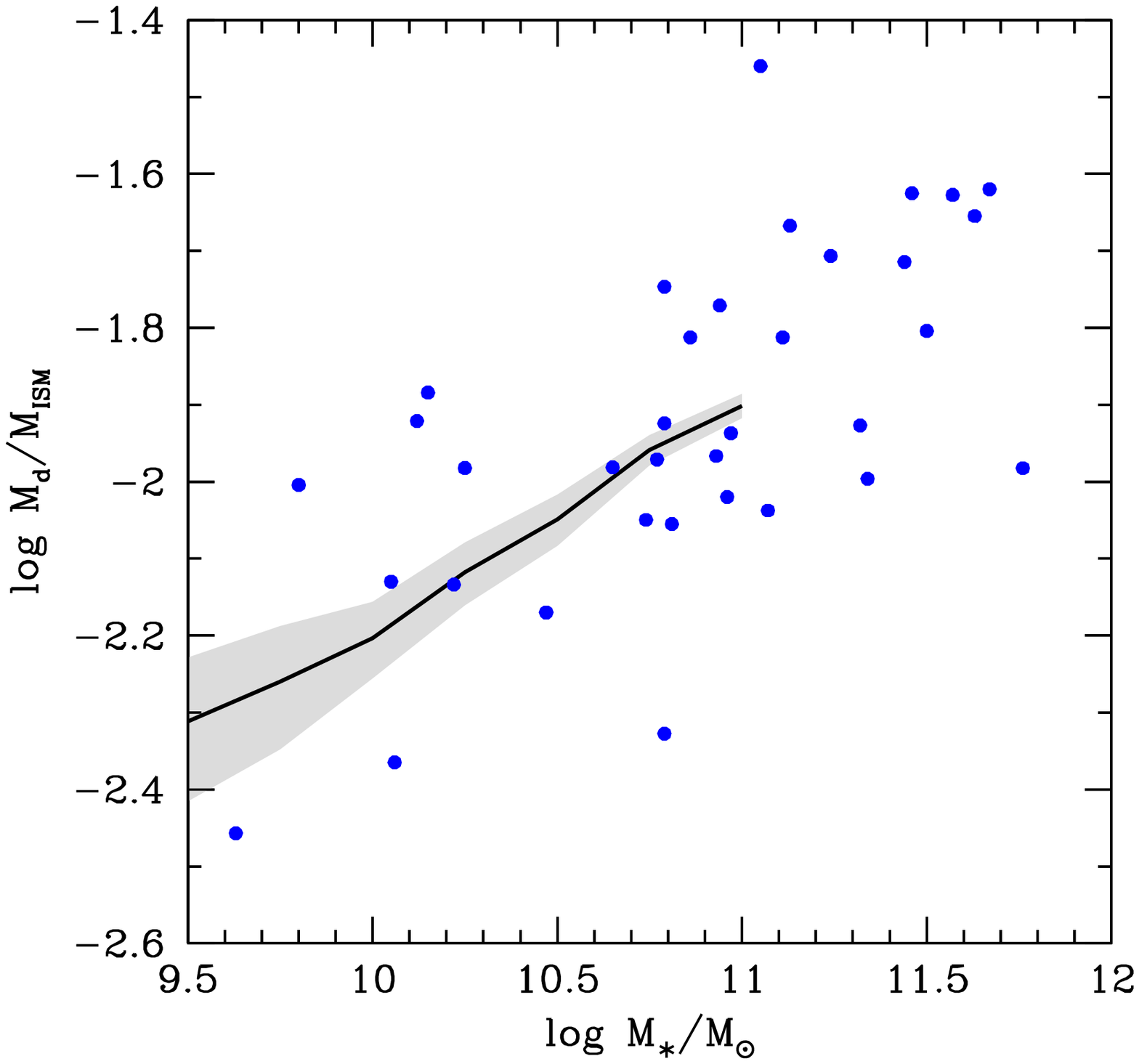}
\hspace{-0.6cm}
\includegraphics [width=6.4cm]{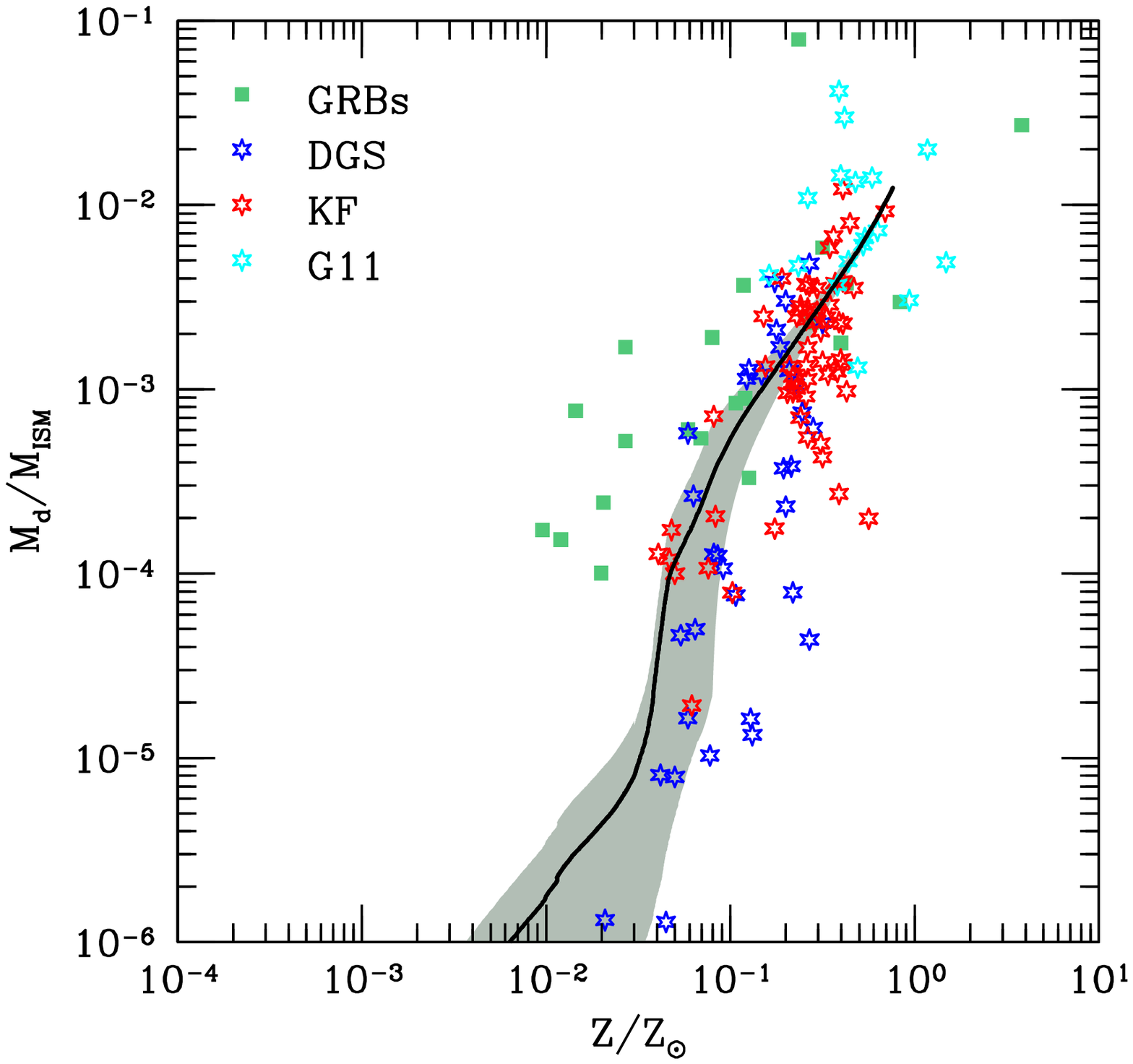}
\vspace{-0.5cm}
\caption{Comparison between the predicted properties of MW progenitors and observations. Solid lines represent averages over 50 independent merger histories of the MW and
shaded region the 1-$\sigma$ dispersion. In the left and middle panels the observational data are taken from Corbelli et al. (2012) with dashed lines indicating the best-fit linear relation for the data. In the right panel, stars refer to the sample analyzed by 
R$\rm \acute{e}$my-Ruyer et al. (2014) and squares indicate the GRB data sample of 
Zafar \& Watson (2013, see text). 
{\it Left panel}:  correlation between the dust mass and the diffuse, molecular and total gas masses (from bottom to top). 
{\it Middle panel}: dust-to-gas mass ratio as a function of the stellar mass.
{\it Right panel}:  dust-to-gas mass ratio as a function of the gas metallicity.}
\label{fig:mwprog}
\end{figure*}
%%%%%%%%%%%%%%%%%%%%%%%%%%%%%%%%%%%%%%%%%%%%%%%%%%%%%%%%%

\section{The global properties of the Milky Way}
\label{sec:calibration}

Since the contribution
of Pop III stars to the total SFR is only relevant at high redshifts, the model predictions for global MW properties and for its progenitors are fairly independent of the adopted Pop III IMF and Pop III/II transition criteria. Fig.~\ref{fig:calib} shows the result obtained for a model where the Pop III/II transition 
occurs when the
dust-to-gas ratio in the dense phase of the ISM exceeds the critical value of $\rm \mathcal{D}_{cr} = 4.4 \times 10^{-9}$ and Pop III stars form according to a Larson
IMF with masses in the range $\rm 10 - 140~M_{\odot}$ (model f in Table~\ref{table:models2}). Hereafter we refer to this model as our reference model.
Its is clear that metal and dust enrichment prevents Pop III stars from forming
at z $< 12$ and that the star formation history is completely dominated by Pop II stars, with a final SFR in good agreement with the observed value.

The model presented in the previous Section relies on a number of free parameters, such as the star formation efficiency,  $\rm \epsilon_\ast$, the SN-driven wind efficiency $\rm \epsilon_w$, and the
parameters  $\textit{f}$  and $\rm \alpha_{MC}$ that control the 2-phase structure of the ISM. In addition, we have to define the typical density and temperature that characterize the dense phase
of the ISM and that define the dust accretion timescale through Eq.~\ref{eq:tauacc}.
Our approach is to constrain their values comparing the model predictions with some observed properties of the MW, such as the current SFR, the 
existing $\rm M_{ISM}$ and $\rm M_\ast$, $\rm M_{ISM}^{MC}$, Z, 
$\rm M_{d}$ inferred from observations targeting regions of larger column densities (that we associated to the MC phase) and regions of lower column density (that we 
associate to the diffuse phase). In Table \ref{tab:masses}, we list the observational data that we have collected from the literature. Fig.~\ref{fig:calib} shows the results
obtained for the selected set of reference parameter values listed in Table~\ref{table:models1}. Each line plotted in the figures has 
been obtained averaging over 50 independent merger trees of the MW, with shaded regions (where present) showing the 1-$\sigma$ dispersion. In the left panel of Fig.\ref{fig:calib},
we show the evolution of the total gas, stellar, metal and dust masses in the ISM of the MW, where at z $> 0$ the quantities have been obtained summing over all the individual MW progenitors
present at the corresponding redshift. The set of reference parameters selected leads to a final MW stellar and gas masses of $\rm M_\ast = 7 \times 10^{10} M_\odot$ and 
$\rm M_{ISM} = 9 \times 10^{9} M_{\odot}$, in good agreement with the observational data (see Table~\ref{tab:masses}). Similarly good is the agreement between the 
predicted final masses of metals, $\rm M_{Z} = 2.7 \times 10^8 M_\odot$, and dust $\rm M_{d} = 1.1 \times 10^8 M_\odot$, and the observational value; note that the data point representing the total mass in metals has been derived from the observed metallicity and gas mass, while the data point of the total dust mass
has been computed as the sum of the observed dust masses of dense and diffuse phases (see Table~\ref{tab:masses}). 
The predicted redshift evolution of the gas, metal and dust masses in the two phases of the ISM are shown in the middle panel of Fig.~\ref{fig:calib}. It is clear that
the selected values for the $\textit{f}$  and $\rm \alpha_{MC}$ parameters lead to final masses of cold dense gas and dust masses in the two phase 
in very good agreement with the observations. Note that the metal mass shown in this figure represents the total mass in heavy elements, summing metals
in the gas-phase and in dust grains. Hence, we find that in the cold dense phase dust accretion is maximally efficient, i.e. at the final redshift all the metals
in the gas-phase are depleted on dust (the green-dashed and red-dashed curves overlap). 
In the right panel of Fig.~\ref{fig:calib}, we also show the predicted redshift evolution of the SFR, separating the contribution of Pop III and Pop II stars to the total star formation rate. 
As already described in Sec.~\ref{sec:new}, we assume that only gas in the diffuse phase can be completely ejected out of the halo, as a consequence of mechanical feedback from SNe. 
Hence, while the gas in the dense molecular phase can be heated and dispersed in the diffuse phase, it is not ejected out of the halo. We explored different mass-loading prescriptions of SN-driven outflows (e.g. assuming that the ejected gas is composed by different mass fractions of diffuse and dense gas), and we find that the results of the model are very similar once a re-calibration of the parameter $\rm \alpha_{MC}$ is made so as to reproduce the existing molecular mass in the MW (see Table~\ref{tab:masses}).
%While the model predictions discussed so far are fairly independent of the adopted Pop III IMF and Pop III/II transition criteria, this is not the case for the contribution
%of Pop III stars to the total star formation rate at high redshift. The figure shows the result obtained for model (d), where the Pop III/II transition occurs when the
%dust-to-gas ratio in the dense phase of the ISM exceeds the critical value of $\rm \mathcal{D}_{cr} = 4.4 \times 10^{-9}$ and Pop III stars form according to a Larson
%IMF with masses in the range $\rm 10 - 300~M_{\odot}$ (see Table~\ref{table:models}). Its is clear that metal and dust enrichment prevents Pop III stars from forming
%at $z < 12$ and that the star formation history is completely dominated by Pop II stars, with a final SFR in good agreement with the observed value.

%%%%%%%%%%%%%%%%%%%%%%%%%%%%%%%%%%%%%%%%%%%%%%%%%%%%%%%%%
\begin{table*}
 \caption{In this table we identify the models that we have investigated varying the  Pop III stellar IMF, the adopted SN yields, and the Pop III/II transition criterium.}
\begin{center}
\begin{tabular}{|c|c c c c c c c c c|}
\hline 
\rm
model name & a& b &c&d&e&f&g&h&i\\
Pop III IMF &PISN+cc&PISN+faint&faint&PISN+cc&PISN+faint&faint&PISN+cc&PISN+faint&faint\\
Pop III/II transition & $\rm Z_{cr}$ & $\rm Z_{cr}$ & $\rm Z_{cr}$ & $\rm \mathcal{D}_{cr}$& $\rm \mathcal{D}_{cr}$& $\rm \mathcal{D}_{cr}$ & $\rm D_{trans,cr}$ & $\rm D_{trans,cr}$ & $\rm D_{trans,cr}$\\
\hline
\end{tabular}
\label{table:models2}
\end{center}
\end{table*}
%%%%%%%%%%%%%%%%%%%%%%%%%%%%%%%%%%%%%%%%%%%%%%%%%%%%%%%%%

\section{Properties of the Milky Way progenitors}
\label{sec:dusty}

Once the free parameters of the model are constrained, we can analyse the properties of the MW progenitors.
The idea is to test if the relations between the dust/metal/gas masses that we predict for our reference model at different evolutionary stages are consistent with the observed scaling relations. Most of the available data refer to galaxies of different masses in the Local Universe. Hence, a meaningful comparison between MW progenitors at different redshifts and present-day galaxies requires that sources with comparable stellar/dust/gas masses are identified. Hence, we apply different mass selection criteria to the MW progenitors depending on the specific data sample that we consider, as it will be discussed below.\\

The first sample that we compare with has been recently analyzed by Corbelli et al. (2012). It refers to galaxies in the Virgo Cluster that have been mapped by the Herschel satellite (HeViCS, Davies et al. 2010), observed at 21 cm and in the CO (1-0) line. The sample comprises 35 late-type galaxies in a dynamical range $\rm Log\, (M_{gas}/M_\odot) = 8 - 10$. Selecting MW progenitors in the same dynamical range at all redshifts, in the left and middle panels of Fig.~\ref{fig:mwprog} we compare the model predictions with observations. The behaviour of the simulated progenitors is globally consistent with observed data, with a slope close to the proposed linear regression (dashed lines). The correlation between the dust mass and total gas mass  shows that MW progenitors are generally more gas-rich than typical galaxies in the observed sample, probably as a consequence of the earlier evolutionary stages of the model galaxies, which lie in the redshift range $\rm 0 < z < 4$. It is important to note that galaxies in the Corbelli et al. (2012) sample
have been classified to be HI-deficient, probably due to galaxy interactions in the cluster environment where they are embedded. Indeed, the model galaxies show a correlation between the dust and molecular gas masses very close to the observations, while, for a given dust mass, the predicted HI masses are generally a factor $\sim 2$ larger than observed.
This is further confirmed by looking at the middle panel of Fig.~\ref{fig:mwprog} where we represent - for the same sample galaxies - the correlation between the dust-to-gas mass ratio and the stellar mass. Model predictions show that MW progenitors are characterized by a  dust-to-gas ratio that grows with the stellar mass, consistent with the observations.
% however, for a given stellar mass, the model galaxies have smaller dust-to-gas ratios with respect to the observed ones.
%Note that in the model galaxies the correlation tends to steepen for the largest stellar masses, as a consequence  of the larger gas consumption into stars.
Yet, predicted stellar masses are smaller than $\rm M_\ast \le 7 \times 10^{10} M_{\odot}$, which is the final stellar mass predicted for the MW at z = 0.
Hence, we conclude that although the Corbelli et al. (2012) sample comprises late-type galaxies in the Virgo clusters, 30\% of which have stellar masses larger than the MW progenitors investigated by the 
present model, the general scaling relations that have been inferred from the observations are well reproduced, with differences that can be ascribed to environmental or evolutionary effects.

Additional constraints on the model can be obtained by comparing the predicted correlation between the dust-to-gas ratio and the gas metallicity. Recently, this correlation has been measured over a 
large metallicity range using different samples of local galaxies, such as the KINGFISH survey (KF, Kennicutt et al. 2011), the sample collected by Galametz et al. (2011, G11), and the Dwarf Galaxy Sample (DGS, Madden et al. 2013). In the right panel of Fig. \ref{fig:mwprog}, we show the data points from these three surveys as recently collected by R$\rm \acute{e}$my-Ruyer et al. (2014). For a comparison, we also show in the same plot the data collected by Zafar \& Watson (2013) over a wide redshift range  $\rm 0.1 < z < 6.3$ using GRB afterglows. Note that while the dust masses in local samples have been inferred from the FIR/sub-mm emission, GRB afterglows allow to investigate the evolution of dust content in their host galaxies using extinction data. In particular, following Kuo et al. (2013), for each GRB we consider the best-fit extinction curve (LMC, SMC or MW-like) and compute the dust-to-gas ratio from the observed optical depth and gas column density, using the corresponding correction factor. The observed data points indicate that the dust-to-gas ratio grows with the metallicity but with a large scatter, particularly at the lowest metallicities. %A similar behaviour is found for the simulated MW progenitors where, at $\rm Z > 0.02\, Z_{\odot}$, the accretion timescale is $\rm \tau_{acc} < 100\, Myr$ and the 
%correlation between the dust-to-gas ratio and the metallicity is mostly driven by grain growth in dense molecular clouds (Kuo et al. 2013; Asano et al. 2013; Zhukovska 2014).
A similar behaviour is found for the simulated MW progenitors. At low metallicities, dust and metal enrichment proceed proportionally to the stellar yields. 
When $\rm Z  \sim 0.02\, Z_{\odot}$, the accretion timescale is $\rm \tau_{acc} < 100\, Myr$, gas-phase metals accretion on dust grains becomes efficient, steepening the
correlation between the dust-to-gas ratio and the metallicity, which is mostly driven by grain growth in dense molecular clouds  (Kuo et al. 2013; Asano et al. 2013; Zhukovska 2014). 
Once all gas-phase metals are accreted, a saturation condition is reached and the simulated behaviour is driven again by the corresponding stellar yields.

\section{Stellar Archaeology}
\label{sec:arch}

%In this Section, we apply the code \textsc{gamete} presented in previous Sections to Stellar Archaeology. Our main
%aim here is to test whether the observed low-metallicity tail of the MDF and the metallicity distribution of CEMP stars
%can provide useful constraints on the IMF of Pop III stars and the physical processes responsible for the Pop III//II transition.
Hence, we have identified 9 different models that we have tested against the observations: these models are characterized
by a common set of parameters (calibrated to match the observed properties of the MW, as discussed in Section 
\ref{sec:calibration}) but differ by the adopted Pop III stellar mass range and/or Pop III SN yields, and by the assumed
criterium controlling the Pop III/II transition. In Table~\ref{table:models2}, we 
list the properties of each model, identified by a letter running from (a) to (i).

\subsection{Constraints on the Pop III IMF}
\label{sec:imf}

\begin{center}
\begin{figure*}
\vspace{-1.5cm}
\hspace{-3.5cm}
\includegraphics [width=21.cm]{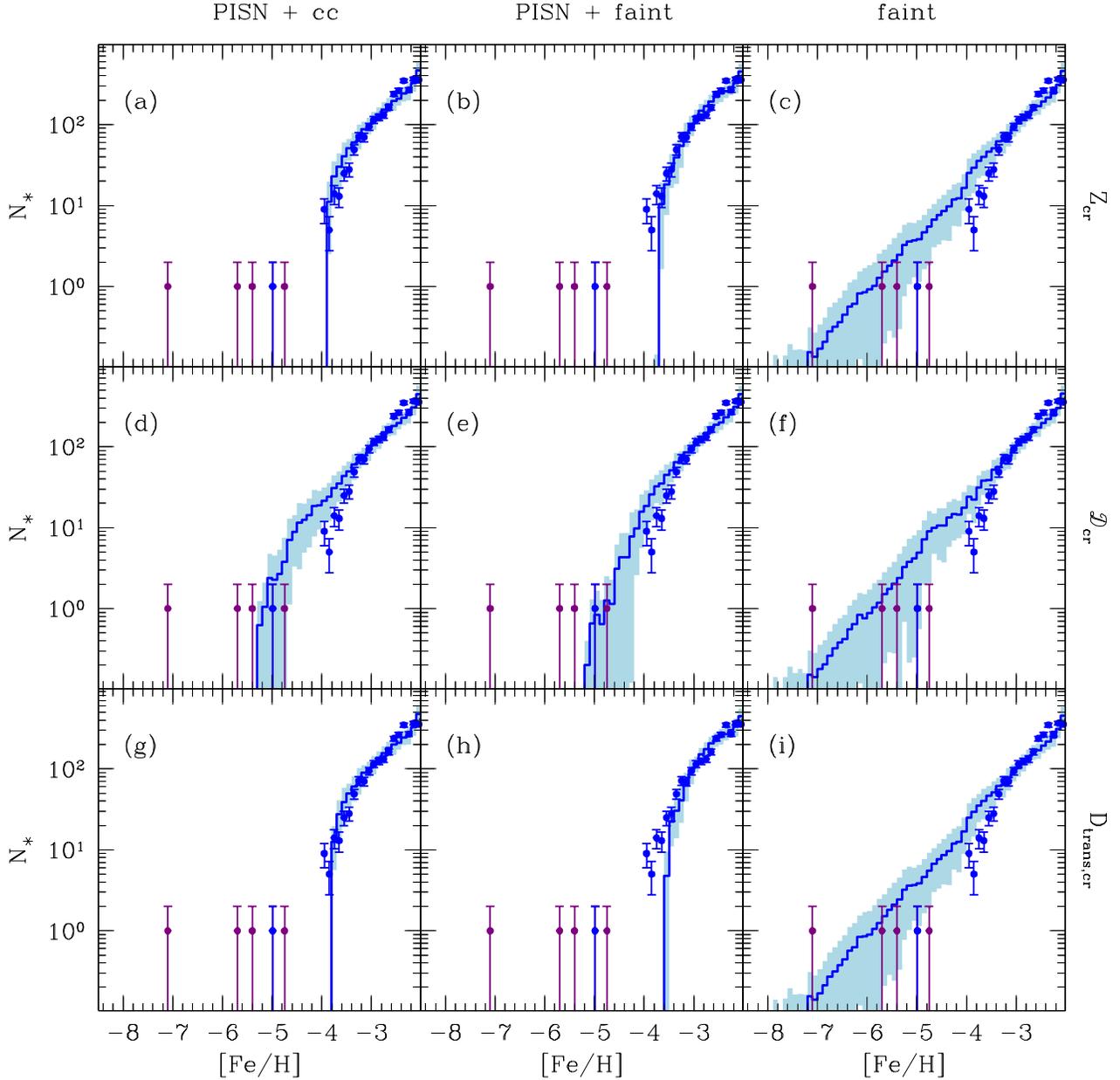}
\hspace{-3.5cm}
\vspace{-2.5cm}
\caption{Simulated MDF averaged over 50 merger trees (solid line) with
1-$\sigma$ dispersion (shaded area) for different Pop III IMF (columns) and for
different transition criteria (rows). Points are the observed MDF as in
Salvadori et al. (2007) with Poissonian error bars, distinguishing between
C-normal (blue) and CEMP (purple) stars. The Pop III/II transition is
parametrized with: a critical metallicity (\textit{first row}) $\rm Z_{cr} =
10^{-3.8} Z_\odot$ (Salvadori \& Ferrara 2012); a critical dust-to-gas ratio (\textit{second row}) $\rm \mathcal{D}_{cr} = 4.4 \times 10^{-9}$ (Schneider et al. 2012a); a critical transition discriminant (\textit{third row}) $\rm D_{trans,cr} = -3.5$ (Frebel \& Norris 2013). The Pop III IMF is taken to extend to: 300 M$_\odot$ with core-collapse SNe (\textit{first column}); 300 M$_\odot$ with faint SNe (\textit{second column}); 140 M$_\odot$ with faint SNe (\textit{third column}). Note that three hyper-iron-poor stars are added with respect to the observed MDF in Salvadori et al. (2007): HE 0557-4840 ([Fe/H] $\approx$ -4.75, Norris et al. 2007); SDSS J102915+172927 ([Fe/H] $\approx$ -4.99, Caffau et al. 2011b; the only C-normal star among the hyper-iron-poor stars); SMSS J031300 ([Fe/H] $\lesssim$ -7.1, Keller et al. 2014).} 
\label{fig:mdf}
\end{figure*}
\end{center}

In Fig.~\ref{fig:mdf}, we show a comparison between the observed low-metallicity tail of the MDF and the theoretical
prediction obtained by each of the 9 models. Note that
we take into account the mass- and metallicity-dependent stellar lifetimes to predict the number of stars that survives down to
$z = 0$. Since at $\rm [Fe/H] \geq -2$ disk stars can contaminate the sample, 
the simulated MDF is truncated above $\rm [Fe/H] = -2$ and it is normalized to the total number of stars observed at $\rm [Fe/H] \leq -2$.
The data points refer to the joint HK/HES sample (Beers \& Christlieb, private communication) that has already been used in
previous theoretical works (Salvadori et al. 2007; 2009). For completeness, we have added to the above sample the 5 hyper-iron-poor
stars currently known: HE0557-4840 ([Fe/H] = -4.75, Norris et al. 2007), SDSSJ102915+1172927 ([Fe/H] = -4.99, Caffau et al. 2011b, 2012),
HE1327-2326 ([Fe/H] = - 5.76, Frebel et al. 2005), HE0107-5240 ([Fe/H] = -5.54, Christlieb et al. 2004), and the recently discovered SMSS J031300 ([Fe/H] $<$ - 7.1, Keller et al. 2014). %Selection biases  have been investigated by Sch{\"o}rck et al. (2009) and Li et al. (2010)
%who concluded that the HES sample is complete at $\rm [Fe/H] \le -3.25$.
%{\bf Note: here add few lines after trying the HES MDF taken from their Table 3 and normalizing the MDF predicted by GAMETE to
%the total number of stars in the sample at $\rm [Fe/H] \le -3.5$.}

%%%%%%%%%%%%%%%%%%%%%%%%%%%%%%%%%%%%%%%%%%%%%%%%%%%%%%%%%
%%% Figure MDF in all cases of IMF and transition criteria
%%%%%%%%%%%%%%%%%%%%%%%%%%%%%%%%%%%%%%%%%%%%%%%%%%%%%%%%%

Let us focus on the first row of Fig. \ref{fig:mdf}, which shows models where the Pop III/II transition is driven by a 
critical metallicity of $\rm Z_{cr} = 10^{-3.8} Z_\odot$ (Salvadori et al. 2007), differing only for 
the adopted Pop III IMF and corresponding yields (see Table~\ref{table:models2}). In model (a) we assume that Pop III 
stars form in the mass range [10-300] M$_\odot$ according to a Larson IMF with $\rm m_{ch} = 20 \,M_\odot$. Hence,
in this model the two mass ranges where Pop III stars contribute to metal and dust enrichment are  
$\rm 10 M_\odot \le m_\ast \le 40\, M_\odot$, where we adopt yields of ordinary core-collapse SNe, 
and $\rm 140 M_\odot \le m_\ast \le 260 M_\odot$,
where the stars explode as PISNe. The resulting MDF is very similar to what was found by previous investigations (see Fig. 6 in Salvadori et al. 2007), 
where a 200 M$_\odot$ single-mass Pop III IMF was adopted. 
This is not surprising because even if we have enlarged the Pop III stellar mass range and relaxed the 
IRA, the IMF-weighted yields are very similar to those of a single 200 M$_\odot$ PISN.  
For this model, the bulk of observed stars at [Fe/H] $\gtrsim -4$ is very well reproduced, 
but no star is predicted with [Fe/H]$< -4$, at odds with the observations. 
Note that 4 out of the 5 stars currently known with $\rm [Fe/H] < -4.5$ are CEMP, hence for these stars [Fe/H] is not a good metallicity
indicator. Although the interpretation of the origin of CEMP stars is complicated by the fact that these stars do not represent a homogeneous
class (see the discussion in Section~\ref{sec:cemp}), 
Pop III SNe in which mixing was minimal and fallback was large have been shown to provide a good match to the
observed abundance pattern (Umeda \& Nomoto 2002; Tominaga et al. 2007;  Joggerst et al. 2009; Kobayashi et al. 2011;
Keller et al. 2014). In these so-called ``faint" SNe, the small iron abundance reflects the fact that 
iron failed to be mixed sufficiently far out to be ejected.

%%%%%%%%%%%%%%%%%%%%%%%%%%%%%%%%%%%%%%%%%%%%%%%%%%%%%%%%%
%%% MDF redshift evolution
\begin{figure*}
\hspace{-0.8cm}
\includegraphics [width=6.4cm]{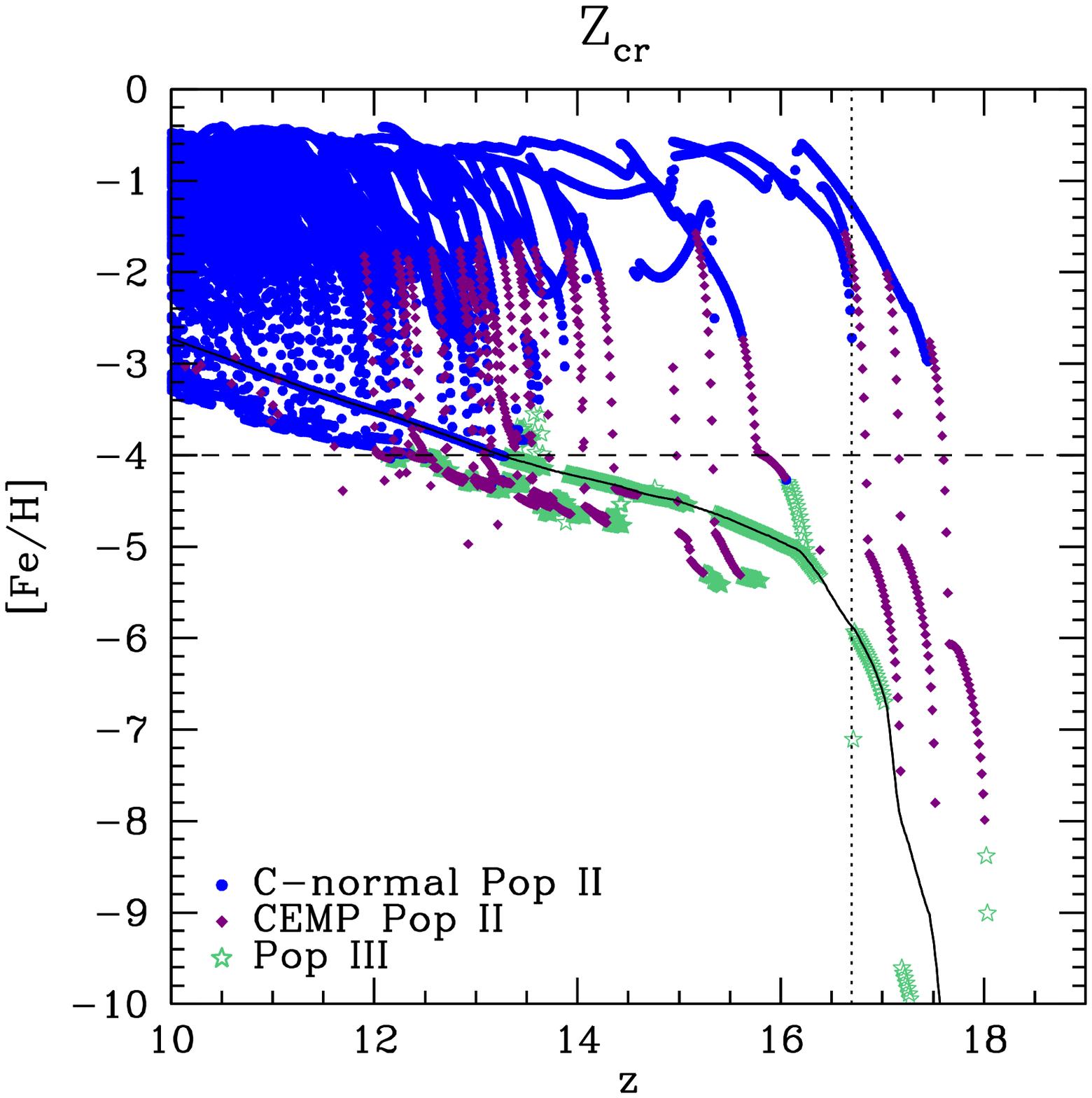}
\hspace{-0.6cm}
\includegraphics [width=6.4cm]{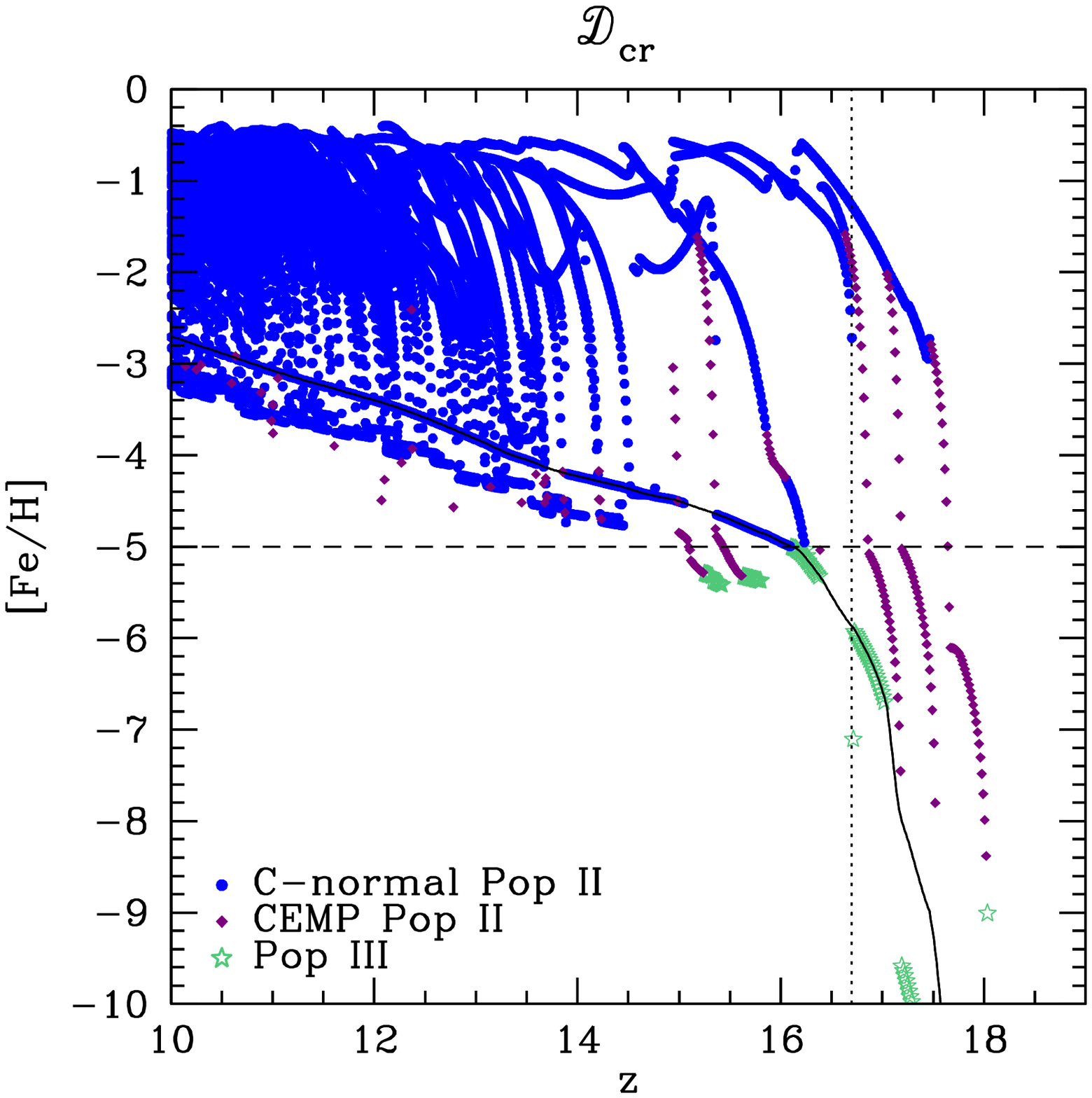}
\hspace{-0.6cm}
\includegraphics [width=6.4cm]{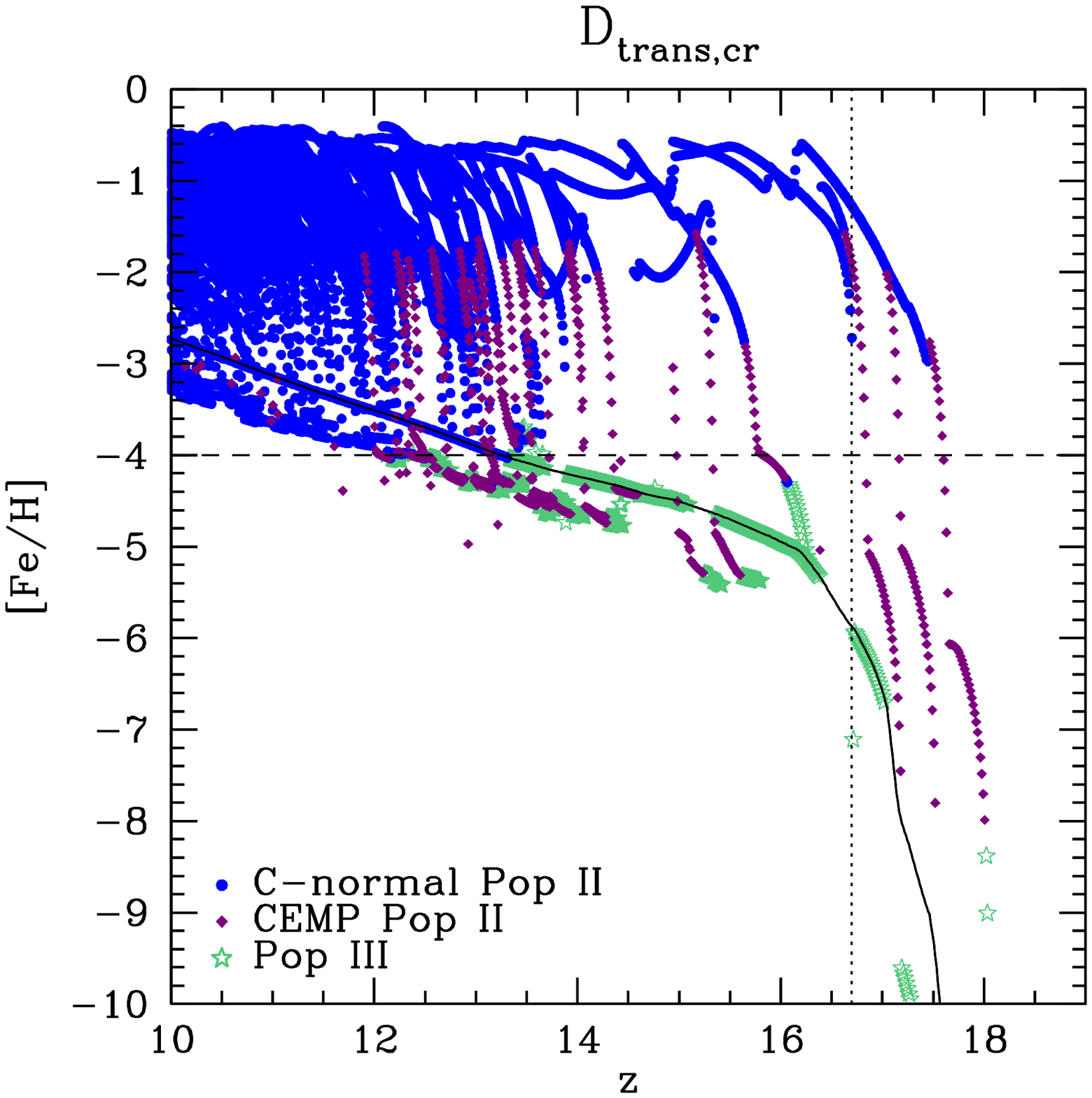}
\vspace{-0.5cm}
\caption{Redshift evolution of the [Fe/H] within MW progenitors for a single merger tree realization for models (c), (f), and (i) (from left to right). Green stars identify 
MW progenitor halos that are forming Pop III stars, while filled purple diamonds and filled blue dots are progenitor halos forming carbon-enhanced and carbon-normal Pop II stars, respectively.
The black solid line represents the evolution of the [Fe/H] of the GM. The vertical dotted lines indicate the redshift at which the GM is characterized by [C/Fe] $>$ 1.0, so that the gas in halos 
virializing from the GM after this redshift are characterized by a C-normal elemental abundance. The horizontal dashed lines represent the [Fe/H] at which the GM exceeds the value of the critical parameter controlling the Pop III/II transition.}
\label{fig:track}
\end{figure*}
%%%%%%%%%%%%%%%%%%%%%%%%%%%%%%%%%%%%%%%%%%%%%%%%%%%%%%%%%

%%%%%%%%%%%%%%%%%%%%%%%%%%%%%%%%%%%%%%%%%%%%%%%%%%%%%%%%%
%%% MDF with CEMP contribution
\begin{figure*}
\hspace{-0.8cm}
\includegraphics [width=6.4cm]{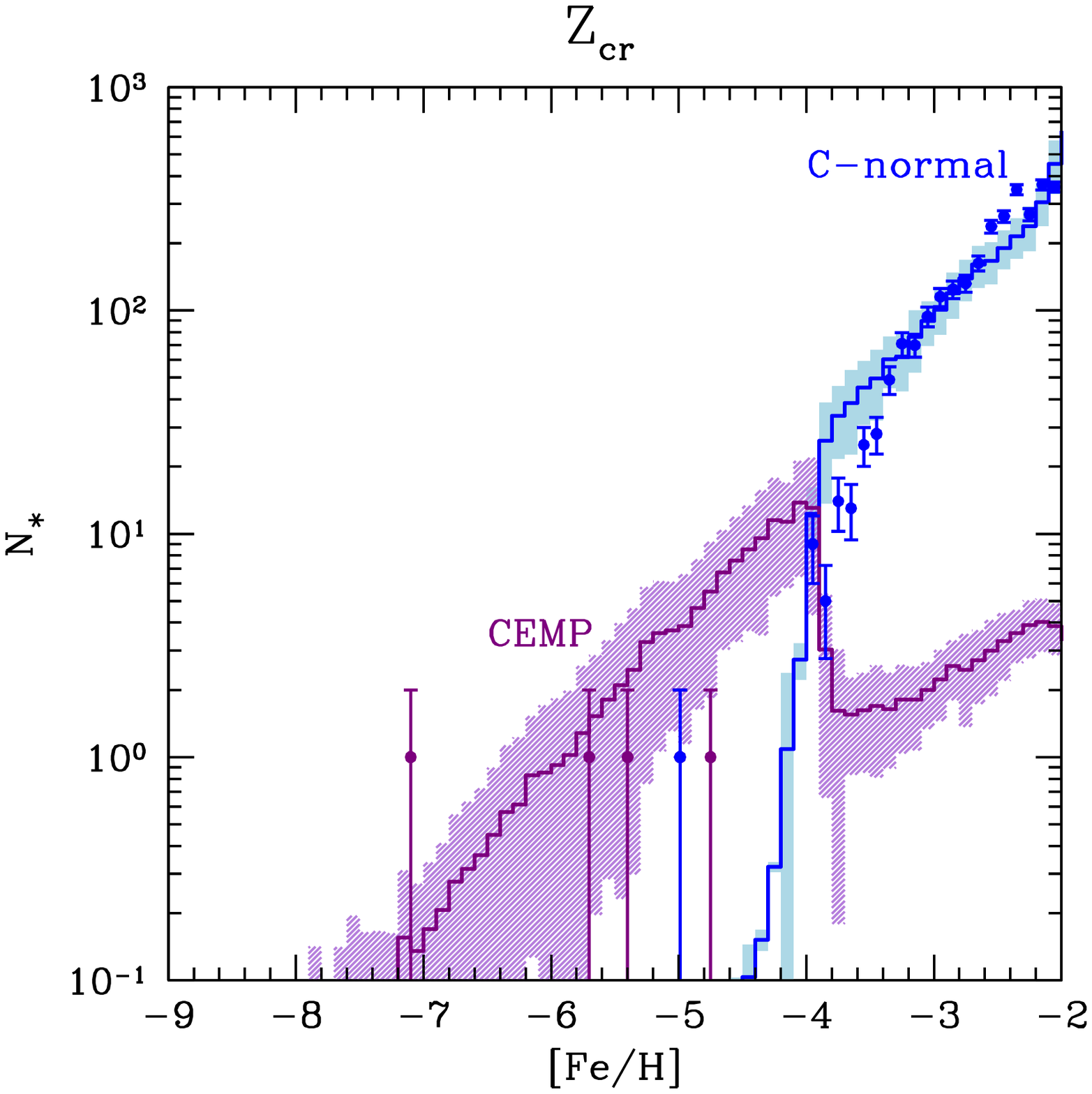}
\hspace{-0.6cm}
\includegraphics [width=6.4cm]{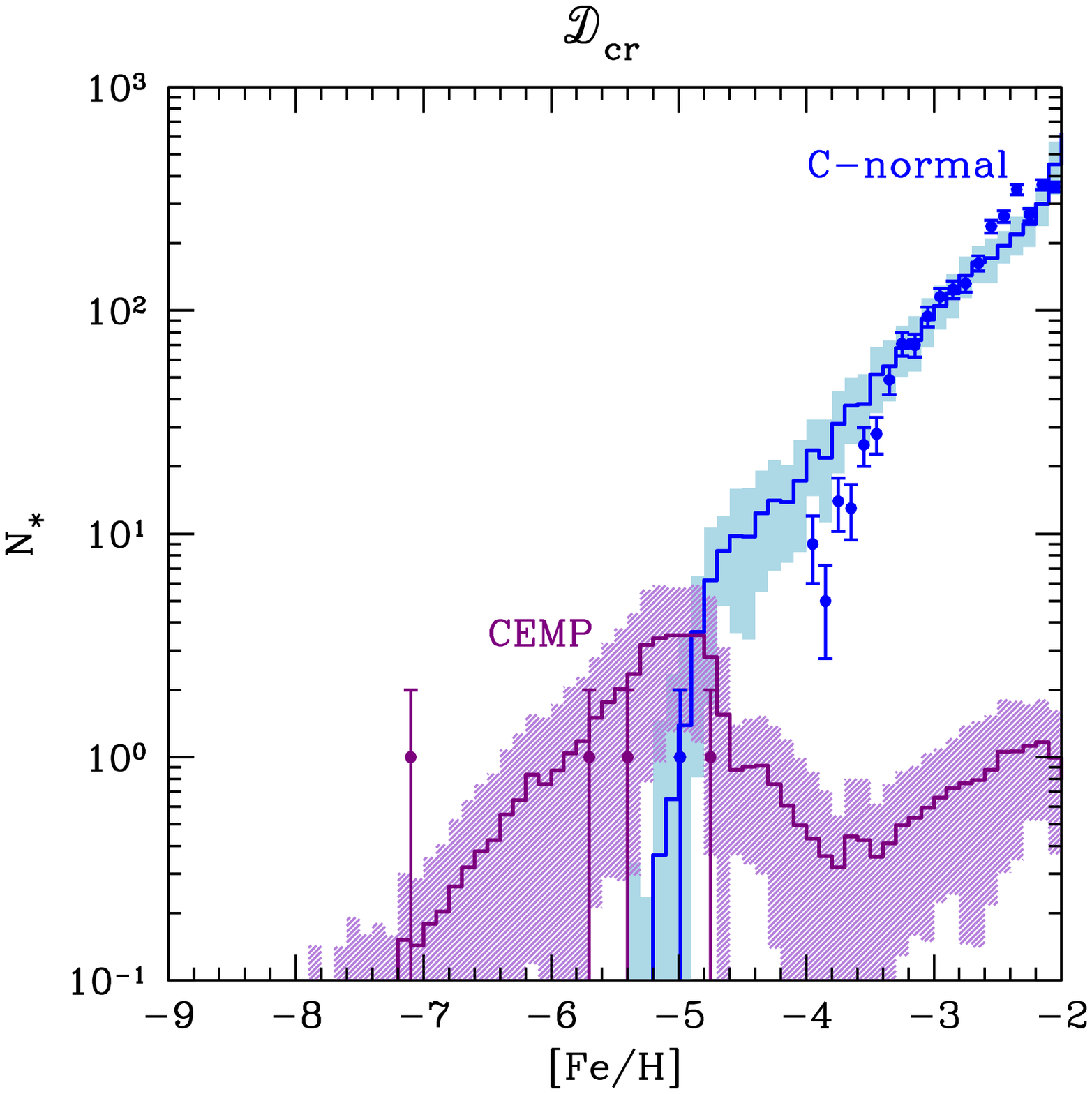}
\hspace{-0.6cm}
\includegraphics [width=6.4cm]{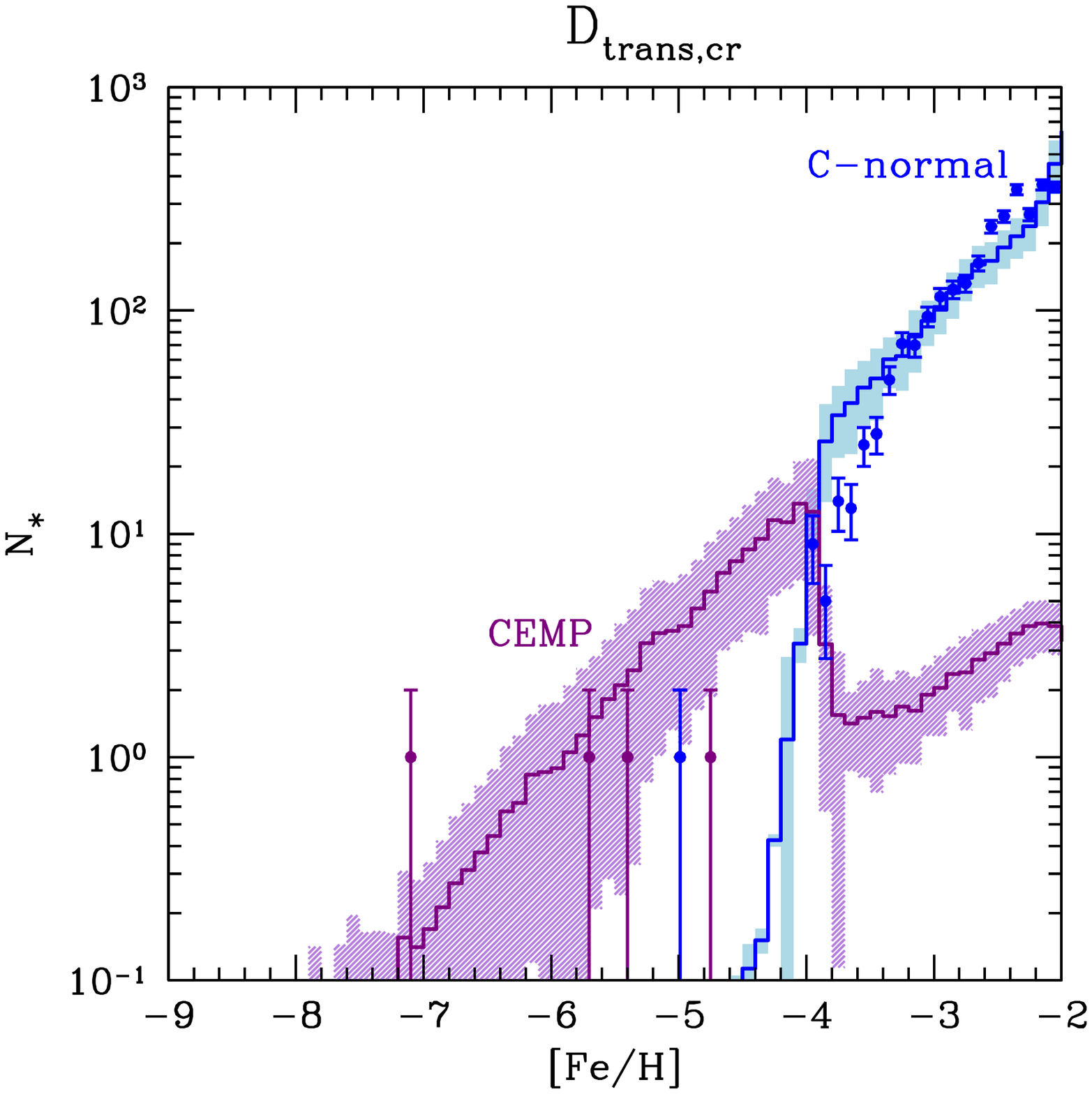}
\vspace{-0.5cm}
\caption{The contribution of C-enhanced stars ( [C/Fe] $\geq$ 1.0, purple line and shaded region) and C-normal stars ([C/Fe] $<$ 1, blue line
and shaded region) to the theoretical MDF for models (c), (f), and (i) (from left to right, see text).}
\label{fig:cemp}
\end{figure*}
%%%%%%%%%%%%%%%%%%%%%%%%%%%%%%%%%%%%%%%%%%%%%%%%%%%%%%%%%

Hence, in model (b) we have adopted
the same Pop III IMF as in model (a), but assuming that Pop III stars with masses in the range 
$\rm 10 M_\odot \le m_\ast \le 40\, M_\odot$ explode as faint core-collapse SNe and contribute to 
metal and dust enrichment with their specific carbon-rich and iron-poor yields, that we take from the recent study by Marassi et al. (2014).
The resulting MDF is shown in the second top panel of Fig.~\ref{fig:mdf} and appears to be almost identical
to that of model (a); in fact, PISN largely dominate the Fe production and - being more massive - are the 
first Pop III stars to enrich the ISM. By the time the first faint SNe explode, the gas is already enriched in Fe and
their contribution to the low-[Fe/H] tail of the MDF becomes negligible. Hence, the lowest 
[Fe/H]-bin populated by stars in the theoretical MDF is related to the PISN iron yield, and the observed
stars at [Fe/H]$ < $-4.5 are still not reproduced by the model.

If we limit the mass range of Pop III stars to $\rm [10-140] \, M_\odot$, excluding the PISN progenitors mass range, we find the results shown in 
the third top panel (model c). Iron enrichment is much slower than previous cases, being dominated by faint SNe. As a result, the tail of the MDF extends 
down to [Fe/H] $\sim$ -8, well into the observed range of the most iron-poor
stars.
%, in agreement to what found by Salvadori \& Ferrara (2012), who assumed 
%Pop III stars to evolve as faint SNe of 25 $\rm M_\odot$ (Kobayashi et al. 2011). 
Although for this model the average MDF agrees with the observations 
at $\rm [Fe/H] \le -4.75$ and at $\rm [Fe/H] \ge -3.5$, at least within the 
$1-\sigma$ scatter, there is a clear excess of stars with $\rm -4.75 < [Fe/H] 
< -3.5$ with respect to the data, which indicate a sharp change of slope in the MDF. The smooth, continuous growth of the simulated MDF for $-8 \le \rm [Fe/H] \le -2$ suggests that additional physical processes, beyond those implemented in the present study, are at work. Most
notably, radiative feedback effects that may prevent the gas in the first mini-halos from cooling, delaying 
or inhibiting star formation (Karlsson 2006; Salvadori \& Ferrara 2012; de Bennassuti et al. in prep). 

%%%%%%%%%%%%%%%%%%%%%%%%%%%%%%%%%%%%%%%%%%%%%%%%%%%%%%%%%
%%% CEMP fraction as function of [Fe/H]
\begin{figure*}
\hspace{-1.0cm}
\includegraphics [width=8.0cm]{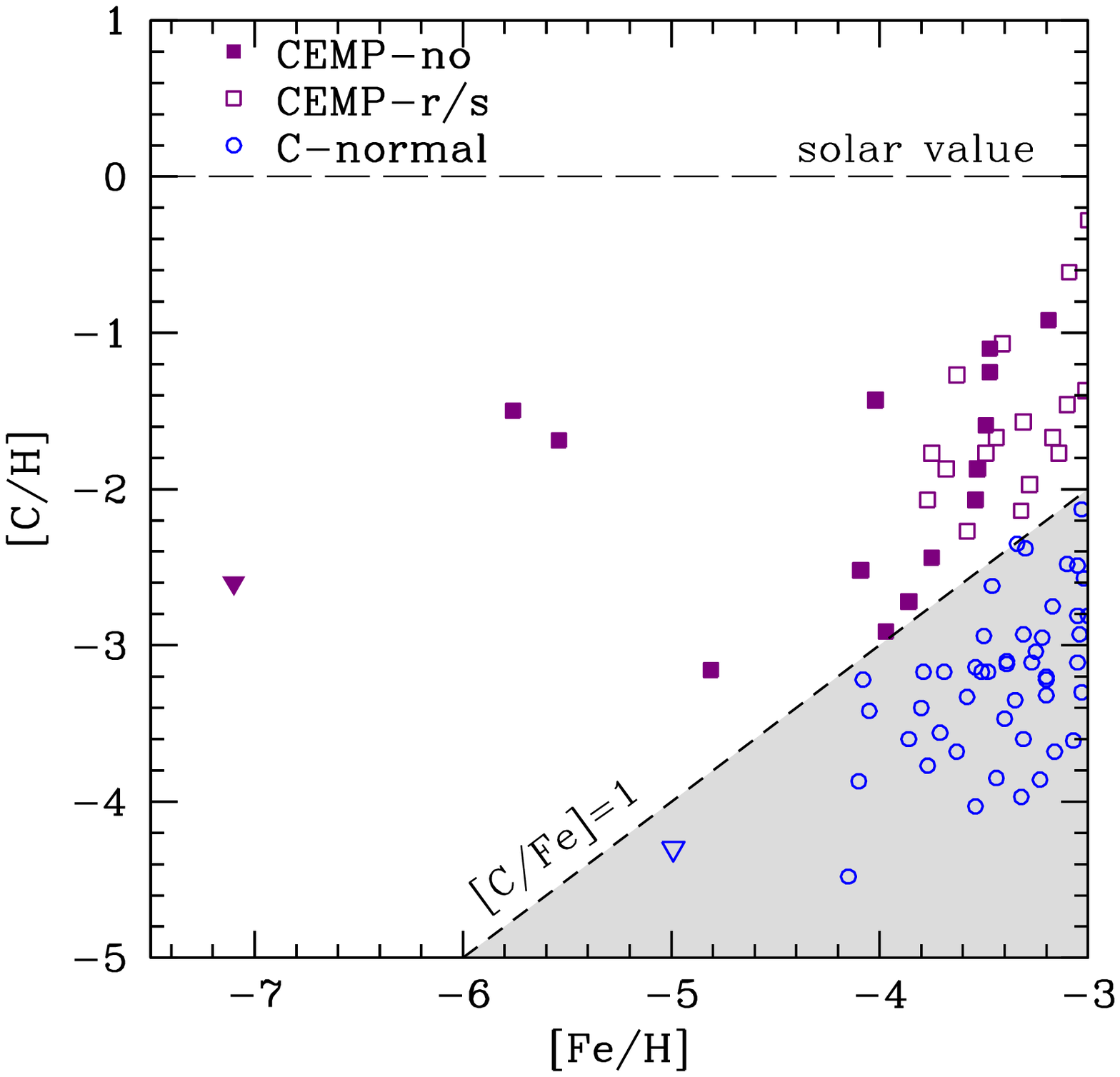}
\includegraphics [width=8.0cm]{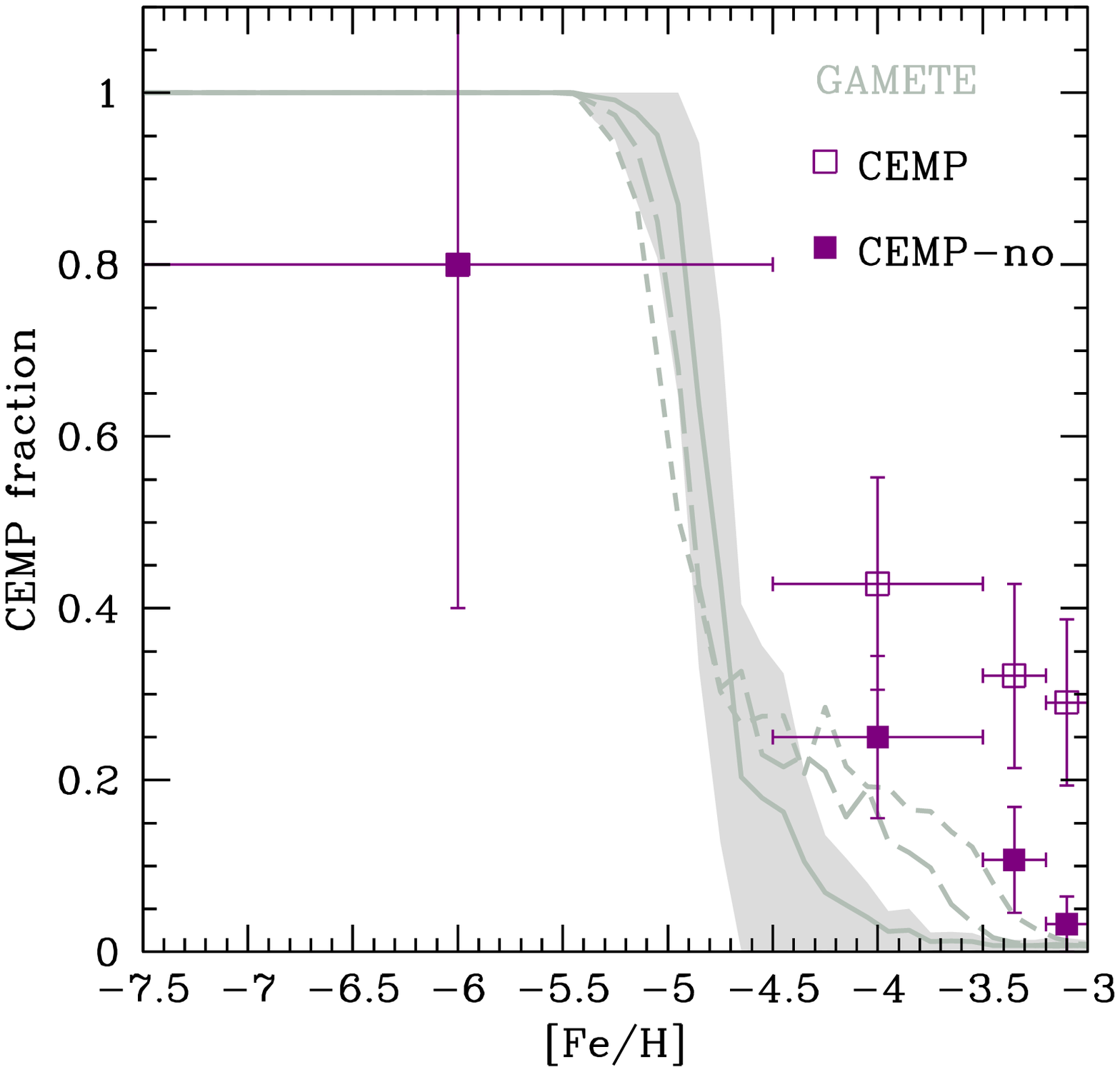}
\vspace{-0.5cm}
\caption{{\it Left panel}: C abundance as a function of [Fe/H] for the C-normal (empty blue points), CEMP-no
(filled purple squares) and CEMP-r/s (empty purple squares) stars taken from the sample of Yong et al. (2013a). We also added to their sample the CEMP-no SMSS J031300 star (purple filled triangle, Keller et al. 2014) and the C-normal SDSSJ102915+1172927 star (blue empty triangle, Caffau et al. 2011b, 2012).
The dashed line corresponding to [C/Fe] = 1 marks the separation between C-normal (grey area) and C-rich stars, following the definition of Beers \& Christlieb (2005). The horizontal dashed line shows the solar
C-abundance. {\it Right panel:} observed fraction of CEMP (empty squares) 
and CEMP-no (filled squares) stars in four different [Fe/H]-bins. Errorbars on the y-axis represent Poissonian 
errors on the number of stars in each bin and on the x-axis reflect the amplitude of the [Fe/H] bin (see text). 
The grey solid line shows the predicted fraction of
CEMP-no stars for the fiducial model (f) averaged over 50 merger trees. The
shaded region shows the 1-$\sigma$ dispersion among different merger histories. 
The long-dashed and dashed lines show the results obtained when 30\% and 50\% 
of the SN type II progenitors are assumed to explode as faint SNe,
respectively.}
\label{fig:cempf}
\end{figure*}
%%%%%%%%%%%%%%%%%%%%%%%%%%%%%%%%%%%%%%%%%%%%%%%%%%%%%%%%%

\subsection{Constraints on the Pop III/II transition}
\label{sec:transition}

The second and third rows of Fig.~\ref{fig:mdf} show - for the same set of Pop III IMF and SN yields discussed above, the predicted
MDF when the Pop III/II transition occurs above a minimum dust-to-gas ratio or transition discriminant. 
Models (d) and (e) show that the condition ${\cal D} \ge {\cal D}_{\rm crit}$ allows to
populate the MDF down to [Fe/H] $\sim -5.4$ even when Pop III stars are assumed to form with masses $\rm [10 - 300] \, M_\odot$
and independently of the adopted core-collapse SN yields. Yet, even in this case the mass range of PISN progenitors has to be excluded 
from the Pop III IMF in order to reproduce the low-[Fe/H] tail down to the lowest observed values (model f). 

Finally, for a given Pop III IMF and SN yields, a Pop III/II transition occurring when $\rm D_{trans} > D_{trans,crit}$
does not lead to appreciable differences in the MDF with respect to models where low-mass stars form when $\rm Z > Z_{cr}$
(compare models g, h, i, with models a, b, c, respectively). In conclusion, faint SNe are needed to reproduce the low-metallicity tail of the MDF.

In Fig.~\ref{fig:track} we show the redshift evolution of the [Fe/H] within MW progenitor halos selected from a single merger-tree realization,
whose results at z = 0 are most similar to the average ones. The three panels refer to models which adopt the same Pop III IMF
( $\rm [10 -140]\, M_\odot$ mass range with faint SNe) but different Pop III/II transition criteria (models c, f, and i, from left to right).
Depending on the gas properties, at each time, individual MW progenitors can either
form Pop III stars (green starred points) or Pop II stars (C-normal or C-enhanced). 
Following the definition of Beers \& Christlieb (2005), metal-poor
stars are classified as CEMP if $\rm [C/Fe] \ge 1$. 
The solid black line represents the [Fe/H] evolution
in the GM. Points along this line identify halos that have virialized from the GM and that form stars for the first time; hence their gas properties
reflect the corresponding properties of the GM at that redshift. At each given redshift, points above the black solid line represent 
halos that have been self-enriched by their stellar populations or that have inherited their [Fe/H] through mergers with highly enriched  
progenitors. Conversely, points below the black solid line represent progenitors whose [Fe/H] has been diluted below the GM value by
mergers with un-polluted halos (that have virialized at an earlier epoch but have not formed stars yet).
The horizontal-dashed and vertical-dotted lines mark two important evolutionary phases: the first one identify the 
[Fe/H] when the critical Pop III/II transition criterium is met in the GM. Hence, halos that virialize after this epoch can only form Pop II stars
(see the transition from starred points to dots along the GM line). This occurs at different epochs in the three models as a consequence of
the different Pop III/II transition criteria (at $z \sim 14$ for models c and i, and at $z \sim 16$ for model f). The nature of the first Pop II
stars that form in newly virialized halos depends on the [C/Fe] content of the GM: the vertical-dotted line marks the redshift at which
[C/Fe] = 1; for all the three models, the first Pop II stars that form in newly virialized halos are always C-normal. 
The typical evolutionary tracks in the first progenitors show that Pop III stars continue to form until the first faint SNe explode and enrich the
ISM above the critical threshold for Pop II star formation. The first Pop II stars are typically C-enhanced, reflecting the yields of faint-SNe. 
Once the first Pop II cc-SNe explode, the [C/Fe] is decreased to $< 1$ and the subsequent generations of Pop II stars are C-normal.

Comparing the three panels, it is evident that when $\rm [Fe/H] < -5 $ the evolutionary tracks do not depend on the  transition criterium,
resulting in a very similar low-[Fe/H] tail of the MDF. The major difference is in the duration of the Pop III star formation epoch, that is shorter 
when a dust-driven transition is adopted. As a result, in model (f) C-normal Pop II stars can form at $\rm  - 5 < [Fe/H] < - 4 $ due to early
enrichment from Pop II cc-SNe. Conversely,  in models (c) and (i) a larger fraction of Pop II stars are C-enhanced due to the 
larger contribution to gas enrichment from faint Pop III SNe. Hence, 
tighter constraints on the Pop III/II transition  can be set by separating the contribution to the 
MDF of C-normal and C-enhanced stars.  The results are shown in Fig.\ref{fig:cemp}. 
It is clear that CEMP
stars dominate the low-[Fe/H] tail of the simulated MDF  for all three models, as a result of the dominant role played by Pop III faint-SN 
in the first metal enrichment. However, it is only in model (f), where Pop II stars from when ${\cal D} \ge {\cal D}_{\rm crit}$,
that the MDF of C-normal stars extends to  $\rm [Fe/H] \sim -5.4$, accounting for the observed data point that
corresponds to SDSS J102915+172927, the only C-normal star at these low-[Fe/H] currently known (Caffau et al. 2011b).
Note that the predicted [C/Fe] for C-normal stars with [Fe/H]$<-4.5$ ranges between 0 and +1, consistent with the observed
properties of SDSS J102915+172927 (Caffau et al. 2011b).
Conversely, the MDF of C-normal stars predicted by models (c) and (i) do not extend below $\rm [Fe/H] \sim -4.5$. Hence, as already
discussed by Caffau et al. (2011b), Schneider et al. (2012b), and Klessen et al. (2012), the formation of
 SDSS J102915+172927 must have followed alternative pathways, beyond that associated to fine-structure line cooling. 

We conclude that the observed MDF at $\rm [Fe/H] \le -2$ 
and the relative fraction of C-normal and C-rich stars at $\rm [Fe/H] \le -4.5$ seem to indicate model (f)
as the best-fit scenario, where the transition is driven by the dust-to-gas ratio,
Pop III stars preferentially form in the mass range $[10 - 140] \rm \, M_\odot$ and explode as faint SNe. The 
contribution of PISN to the first enrichment, if present, must have been largely sub-dominant. Finally, the formation
of the first low-mass stars must have been triggered by processes occurring at metallicities below those normally
assumed to enable efficient fine-structure line cooling, such as by dust cooling and fragmentation. 
Other models can be excluded because they do not predict the existence of stars and/or C-normal stars below 
$\rm [Fe/H] < -4$, at odds with observations.

\subsection{Carbon enhancement}
\label{sec:cemp}

Spectroscopic studies of metal poor stars in the HK, HES and more recently SDSS samples have 
convincingly shown that the frequency of CEMP stars grows with decreasing [Fe/H], being $\sim 20\%$
at $\rm [Fe/H] < -2$ and as large as $80 \%$ at $\rm [Fe/H] < -4$ (Yong et al. 2013b; Spite et al. 2013; Norris et al. 2013;
Lee et al. 2013). However, the increasing frequency of CEMP stars with declining [Fe/H] is not equally important
for the different sub-classes which characterize CEMP stars. Generally, CEMP stars
are classified as CEMP-s/rs and CEMP-no (Beers \& Christlieb 2005) depending on the 
over-abundant presence of s or rs-process elements (such as Ba, Sr, Eu) or no-overabundance of 
neutron-capture elements. It has been suggested
that these differences may provide clues to the nature of their likely progenitors. In particular, 
CEMP-s and CEMP-rs have abundance patterns that suggest mass-transfer from an AGB companion
in a binary systems (Suda et al. 2004; Masseron et al. 2010). Indeed, long-term radial-velocity 
observations show that a large percentage of CEMP-s/rs stars are binaries (Lucatello et al. 2005). 
However, the abundance pattern of the CEMP-no stars is difficult
to explain as caused by a mass transfer from an AGB companion and the nature of
their progenitors is still largely debated. Using a homogeneous chemical
analysis of 190 metal-poor stars Yong et al. (2013a) and Norris et al. (2013) have recently investigated a sample of CEMP-no stars with $\rm [Fe/H] < -3$. 
They conclude that the observed chemical abundances are best explained by models which invoke the nucleosynthesis of Pop III SNe with mixing and 
fallback, the so-called ``faint" SNe that we have discussed in Sec.~\ref{sec:imf}. In addition, observations show that the relative fraction of CEMP-no 
and CEMP-r/rs stars depends on [Fe/H], with CEMP-no stars representing the 
large majority ($\sim 90 \%$) of CEMP stars at $\rm [Fe/H] < - 3$ (Aoki et 
al. 2007; Masseron et al. 2010).

%%%%%%%%%%%%%%%%%%%%%%%%%%%%%%%%%%%%%%%%%%%%%%%%%%%%%%%%%
%%% MDF redshift evolution
\begin{figure*}
\hspace{-0.8cm}
\includegraphics [width=6.4cm]{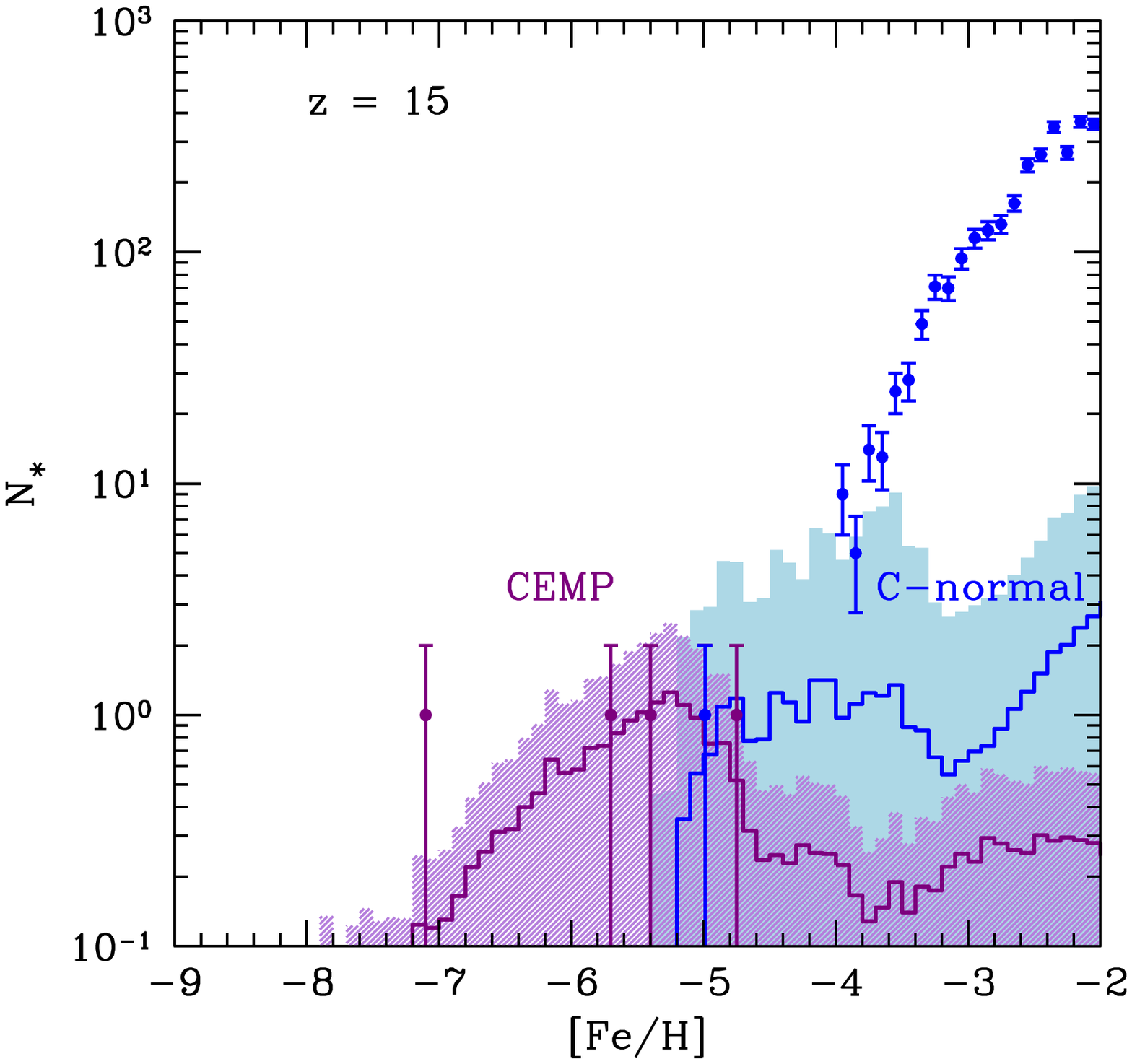}
\hspace{-0.6cm}
\includegraphics [width=6.4cm]{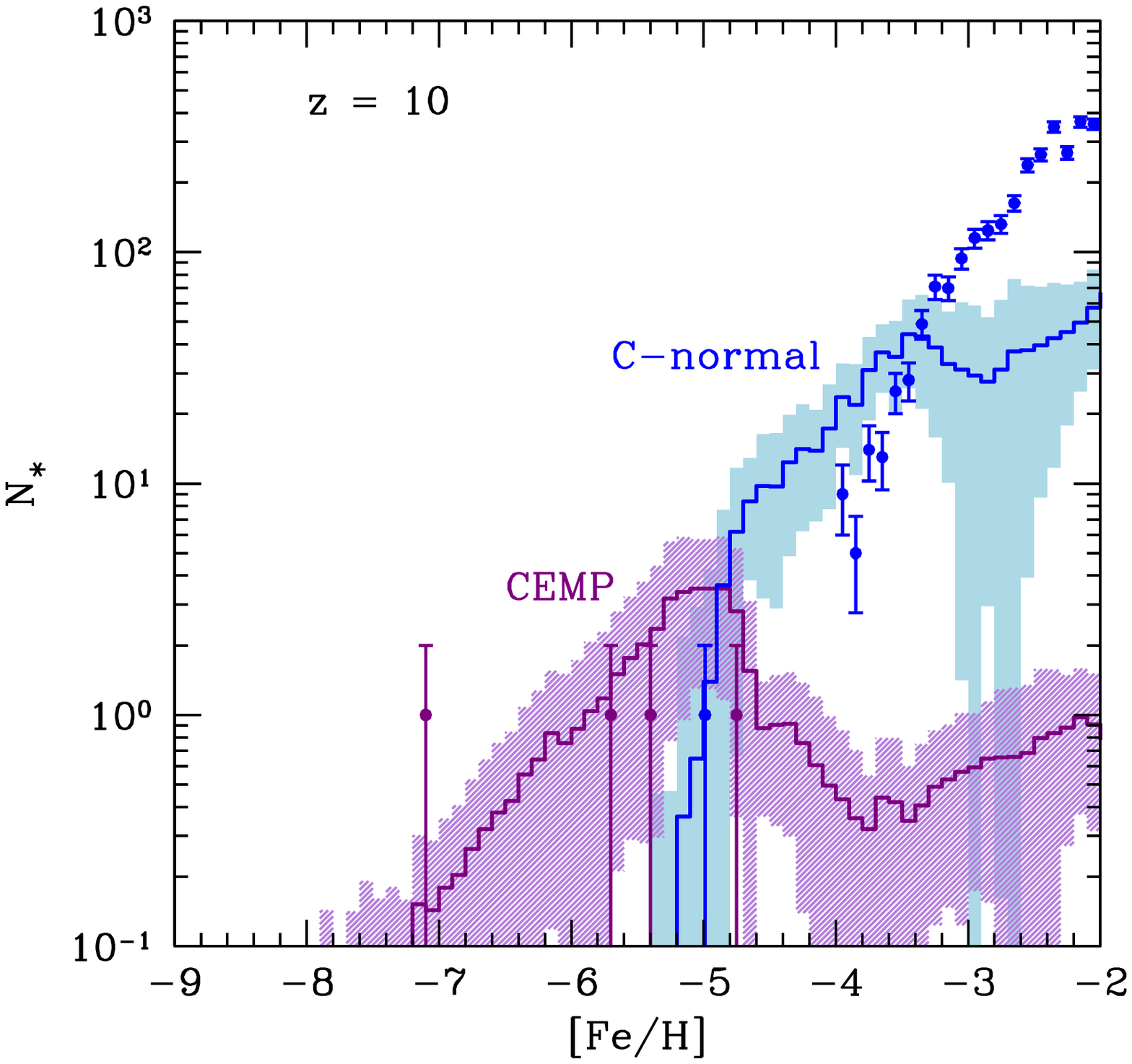}
\hspace{-0.6cm}
\includegraphics [width=6.4cm]{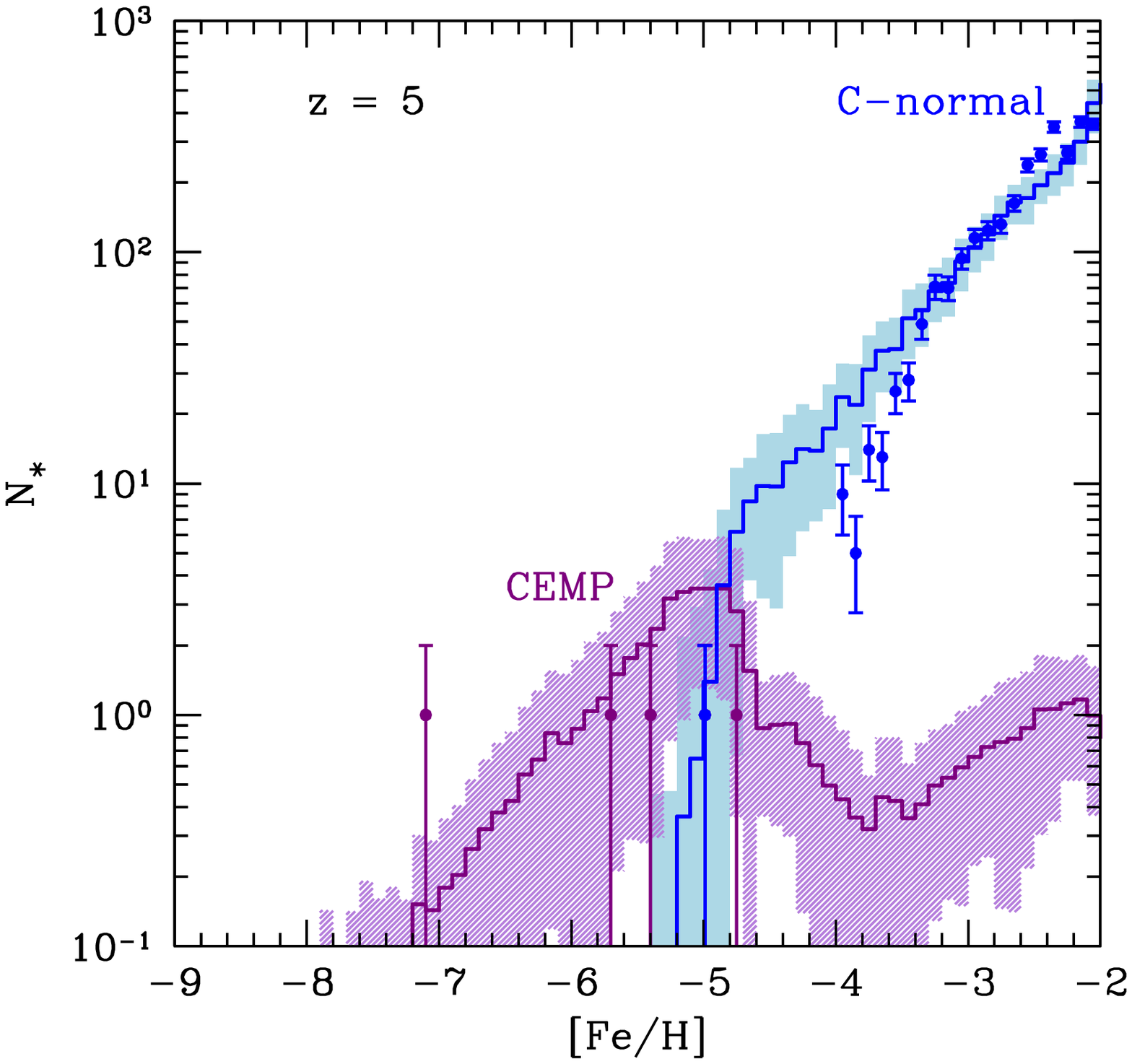}
\vspace{-0.5cm}
\caption{Redshift evolution of the MDF for the fiducial model (f), separating the contribution of CEMP and C-normal stars (z = 15, 10, 5 from left to right).}
\label{fig:mdfredshift}
\end{figure*}
%%%%%%%%%%%%%%%%%%%%%%%%%%%%%%%%%%%%%%%%%%%%%%%%%%%%%%%%%

In the left panel of Fig.~\ref{fig:cempf} we show [C/H] as a function of [Fe/H] for 86 stars with $\rm [Fe/H] < -3$
out of the full sample of 190 stars analyzed by Yong et al. (2013a, see their Table 1), to which we have added 
SDSSJ102915+1172927 (Caffau et al. 2011b) and SMSS J031300 (Keller et al. 2014), that were not
present in the original sample. Following Norris et al. (2013), we have divided the sample into CEMP-no and 
CEMP-r/s stars, where the latter class includes CEMP-s, CEMP-r, and CEMP-rs stars. The grey shaded area
marks the region of the plane where the stars are classified as C-normal ($\rm [C/Fe] < 1$, according to our
definition). It is clear from the figure that CEMP-no stars represent the dominant population of CEMP stars at 
$\rm [Fe/H] < -4.5$, but that the number of CEMP-r/s stars grows with [Fe/H] and become the dominant class
of CEMP stars at $\rm [Fe/H] > -3.5$. This is clearly seen in the right panel
of the same figure, where we plot the observed fraction of CEMP stars (number of CEMP stars divided by the total number of stars) and of CEMP-no stars 
for four different [Fe/H] bins. To be consistent with Yong et al. (2013b), we have considered a single bin for the
5 hyper-iron-poor stars with $\rm [Fe/H] < -4.5$ and then three bins in the $\rm -4.5 \le [Fe/H] \le -3$ range, chosen to contain
an almost equal number of stars ($\sim$ 28, 27, and 28 from the lowest to the
highest [Fe/H]-bin).
Empty data points refer to the fraction of CEMP stars and filled data points 
to the fraction of CEMP-no stars only.  We find that the fraction of CEMP 
stars decreases from 0.8 to 0.3, but that the decrease becomes much steeper
when only CEMP-no stars are considered, going from $\sim 80 \%$ to less than
$5\%$. Yet, the predicted fraction of CEMP stars for the fiducial case
(f) shows an even steeper decline. It is important to stress that in our 
model CEMP stars can form only as a result of the enrichment from Pop III 
faint SNe, as we do not consider mass-transfer effects in stellar binaries.
Hence, strictly speaking, we can not predict the existence of CEMP-r/s stars 
and the simulated data must be compared only to the CEMP-no fraction. 
However, while the model is consistent with the data (within 1-$\sigma$) for 
three out of the four [Fe/H]-bins, it under-predicts the fraction of CEMP-no stars at 
$\rm [Fe/H] \ge -4.5$. We will further discuss this point in the next section.
Here we comment on the fact that we have conservatively 
assumed that only Pop III stars can end their life as faint SNe, whereas 
faint SNe have been directly observed in the local Universe, like SN1997D 
(Turatto et al. 1998), which has been considered as a prototype for this 
class of objects, and many more examples (see the compilation in Pastorello 
et al. 2006 and Fraser et al. 2011), including the dimmest sources SN 1999br (Pastorello 2004) and SN 2010id (Gal-Yam et al. 2011).
Additional indication that a population of faint SNe may exist come from the non-detection of SNe
associated with low-redshift long-GRBs (Della Valle et al. 2006). 
It is difficult to directly estimate the fraction of faint SN from current SN data, due to the
bias against dim sources in present samples. Using very local ($\rm \le 10 Mpc$) data,
Horiuchi et al. (2011) have estimated that  the fraction of dim SNe, defined as SNe with an absolute
magnitude of M $>$ -15, ranges between 3\%-20\%. However, they also find that a fraction as large as  30\%-50\%
is required in order to reconcile the observed SN rate with that predicted from the observed cosmic SFR. 
Assuming a similar fraction of Pop II  faint-SNe we can better match the observed data points, as shown in
Fig.\ref{fig:cempf}. 

\section{Discussion}
\label{sec:disc}
Several theoretical models aimed to investigate the early enrichment and star formation
history of the MW have been developed and compared to observations of the most metal-poor 
stars in the Galactic halo and nearby dwarf galaxies. 

Tumlinson (2006) and Salvadori et al. (2007) first applied semi-analytic approaches to 
hierarchical structure formation to follow the chemical enrichment of the Milky Way and 
to put constraints on the properties of the first stars.
Tumlinson (2006) explored different Pop III IMFs and found good agreement with the Galactic 
halo MDF for top-heavy IMFs sharply peaked in the mass range of core-collapse SNe, [12-70]
M$_\odot$, and for $\rm Z_{cr} \sim 10^{-4} Z_\odot$.
Salvadori et al. (2007) investigated the Pop III/II transition by adopting different 
$\rm Z_{cr}$ values and Pop III masses. They showed that the persistent non-detection of 
metal-free stars implies that Pop III stars have masses $\rm m>0.9 \, M_\odot$, or that $\rm Z_{cr}>0$. 
By assuming that Pop III stars explode as PISN, moreover, they found that a value of 
$\rm Z_{cr} \sim 10^{-4} Z_\odot$ is required to correctly reproduce the observed MDF 
at [Fe/H]$>-4$, although the existence of hyper-iron-poor stars cannot be explained.
By using an upgraded version of the model, which includes the physics of mini-haloes
along with an analytical prescription to account for the inhomogeneous reionization
and metal enrichment of the Galactic Medium, Salvadori \& Ferrara (2012) showed that
the low-Fe tail of the Galactic halo MDF shifts towards lower [Fe/H] values when Pop III
stars are assumed to explode as faint SNe of $\rm m = 25 \, M_\odot$ (Kobayashi et al. 2011).
By studying the chemical abundances of very metal-poor damped Ly$\alpha$ absorption
systems in the context of the Milky Way formation, these authors demonstrate that faint
SNe dominate the chemical enrichment at [Fe/H]$< -5$, while at higher [Fe/H] the gas (and
stars) chemical abundances are only partially imprinted by this pristine stellar generation.  

Tumlinson (2007) considered an IMF dependent on the temperature of the Cosmic Microwave 
Background (CMB) as a way to explain the [Fe/H]-dependence of CEMP-s/rs stars, assumed 
to form in stellar binaries. 
Komiya et al. (2009, 2010, 2011) addressed the difference between the hyper-iron-poor stars at [Fe/H] $<$ -4.5 and stars with [Fe/H] $>$ -4.5 in the context of hierarchical galaxy formation: they examine the possibility that hyper-iron-poor stars could originate by low-mass Pop III stars whose surface is polluted by accretion of enriched gas from the halo ISM. They discussed the total number of stars with [Fe/H] $<$ -3 as well as the shape of the MDF and found the best agreement for a high-mass IMF (assumed to be the same for all stellar populations) and accounting for the formation of stellar binaries, without evoking a transition in the stellar IMF. Additionally, they explored different SN yields obtaining good fit to the MDF for all adopted SNe nucleosynthesis model.

Very recently, Cooke \& Madau (2014) have suggested that carbon-enhanced metal poor stars may originate
in mini-halos polluted by low-energy supernovae which produce a super-solar ratio of [C/Fe]. Conversely high-energy SNe that produce a near-solar ratio of [C/Fe] completely unbound the gas in the first mini-halos, thereby suppressing the formation of a second generation of stars.
The recent detection of SDSS J001820.5-093939.2, a truly second-generation 
star at $\rm [Fe/H] \sim - 2.5$ exhibiting the clear imprint of PISNe (Aoki et al. 2014), 
suggests that this is not the case. This observation demonstrates that these
stars do exist but they are extremely rare and they are likely hidden at higher 
[Fe/H], as already suggested by Salvadori et al. (2007).

In this paper we investigate the hierarchical formation of the MW and its chemical enrichment in metals, Fe, C, O. For the first time, we develop and 
apply a chemical evolution model in a 2-phase ISM, to account for grain growth in cold, dense clouds and grain destruction in the hot, diffuse
medium. This allows us to investigate the Pop III/II transition under different physical conditions, including the effect of line-cooling ($\rm Z_{cr}$ or $\rm D_{trans,cr}$),
and dust-grains ($\rm \mathcal{D}_{cr}$).
We explore different IMFs and different SNe yields for Pop III stars. The existence of hyper-iron-poor, carbon-enhanced stars is 
a strong indication that faint SNe (rather than PISNe and cc SNe, which give a too strong Fe enrichment) dominate the nucleosynthetic output
of Pop III stars. 
The assumption of Pop III faint SNe is crucial not only to extend the tail of the MDF down to  $\rm [Fe/H] < -7$ , but also to reproduce the dependence of the 
CEMP fraction on [Fe/H]. 
In the present study, we have not considered a binary origin of CEMP stars. For this reason, we have compared our results with observed samples excluding 
CEMP-s/r/rs stars; yet, as shown in Fig.~\ref{fig:cempf}, the fraction of CEMP stars predicted in our simulated samples tends to under-predict the observed
data at $\rm - 4.5 < [Fe/H] < -3$. As we have discussed in the previous section, the observations could be reproduced if 30\%-50\% of Pop II SNe  are assumed
to be faint, consistent with the observational indication of local SN samples (Horiuchi et al. 2011).
The inclusion of mini-haloes into the model is also expected to increase the number of carbon-enhanced stars, providing a better 
agreement with the data (Salvadori et al. in prep).  However, we should note that the $\rm - 4.5 < [Fe/H] < -3$ range is where
the simulated MDF provides the worst agreement with the observations, showing no sign of the observed change of slope between the roughly 
constant tail at [Fe/H] $< -4 $ and the steep rise at [Fe/H] $> -4$. 
On the contrary, even for the
best-fit model, the simulated MDF is characterized by a single slope in the range $\rm -7 < [Fe/H] < -2$, although with an increasing scatter between different
merger tree realizations at $\rm [Fe/H] < -4$. This is not surprising given that we have assumed that radiative feedback has suppressed 
star formation in mini-halos, which are instead predicted to be the progenitors 
of today lived ultra-faint dwarf galaxies (Salvadori \& Ferrara 2009) and to have played a major 
role in the early metal enrichment and reionization of the Milky Way environment (Salvadori et al. 2014). 
In the present study, instead, stars form only in halos with virial temperatures $\rm T_{vir} \ge 2 \times 10^4~K$ that can cool via Ly-$\alpha$ cooling.
Hence the shape of the MDF at low [Fe/H] mostly reflects the efficiency of star formation and metal enrichment in these more massive progenitors. Yet,
Fig.~\ref{fig:mdfredshift} shows that Pop II stars that populate the bins at $\rm -5 < [Fe/H] < -3$ in the simulated MDF mostly form at redshifts
$15 \le z \le 10$, where we expect radiative feedback to play a major role in self-regulating the efficiency of star formation and metal enrichment in mini-halos
(Salvadori \& Ferrara 2012). A proper analysis of low-metallicity star formation in mini-halos will be the subject of a forthcoming paper (de Bennassuti et al. in prep).

\section{Conclusions}
\label{sec:conc}

In this paper, we have presented a new version of the semi-analytical code \textsc{gamete} and some first applications of this new model to Stellar Archaeology. For the first time, we follow the star formation history and the enrichment in metals and dust 
of the MW galaxy along its hierarchical evolution. Using a 2-phase description of the ISM in each progenitor galaxy of the MW,
we can follow the evolution of the atomic and molecular gas components and the separate evolution of metals and dust grains
in these two phases. In particular, dust grains are destroyed by interstellar SN shocks in the hot diffuse phase, but are shielded
against destruction and grow in mass by accreting gas-phase metals in the dense, molecular phase. We show that:

\begin{itemize}
\item The model is characterized by 4 free parameters (efficiencies of star formation and SN winds and two additional parameters
which regulate the mass transfer between the diffuse and molecular phases) that have been calibrated so as to reproduce 
 the observed properties of the MW at $z=0$.
 %, such as the mass of stars, gas (both atomic and molecular), gas metallicity and dust-to-gas ratios.

\item Using this fiducial set of parameters, the simulated progenitors of the MW follow the scaling relations between the gas (total, dense and diffuse) 
and dust, and between the dust-to-gas ratio and stellar mass inferred from observations of Virgo cluster galaxies (Corbelli et al. 2012); in addition,
it  also predicts a relation between the dust-to-gas ratio and metallicity that is in very good agreement with observations of different samples of local
galaxies (R$\rm \acute{e}$my-Ruyer et al. 2014) and GRB-host galaxies at $0.1 \le z \le 6.3$ (Zafar \& Watson 2013). 

\item When applied to Stellar Archaeology, the simulated MDF is very sensitive to the adopted Pop III IMF and metal yields. Current observations seem to 
require a largely sub-dominant contribution from Pop III PISN to early metal enrichment, favouring Pop III stars with masses in the range 
$\rm [10 - 140] M_\odot$ which explode as faint SNe. 

\item The separate contribution of C-normal and CEMP stars to the MDF is sensitive to the critical conditions that are assumed to control the Pop III/II
transition. This analysis confirms the expectations that C-normal stars with $\rm [Fe/H] < -4.5$, such as  J102915+172927 (Caffau et al. 2011b), can form
only if the transition is driven by dust, when ${\cal D} \ge {\cal D}_{\rm crit} = 4.4 \times 10^{-9}$.

\item Our fiducial model predicts a steep decline of the CEMP fraction with [Fe/H], consistent with the data, but under-predicts the fraction of CEMP
stars in the range $\rm -4.5 \le [Fe/H] \le -3$, suggesting the additional contribution from Pop II faint SNe to C-enrichment. 

\item The change of slope in the low-[Fe/H] tail of the observed MDF can not be explained by chemical and
mechanical feedback effects, and may be due to additional physical processes that regulates the efficiency of star formation
in the first star-forming regions. Most likely, radiative feedback associated to the rising UV-background that inhibits 
star formation by photo-dissociating molecular hydrogen and/or suppressing gas infall in the first mini-halos.
 
\end{itemize}

\section*{Acknowledgments}

We thank the anonymous referee for her/his positive and useful comments and suggestions. The research leading to these results has received funding from the European Research Council under the European Union’s Seventh Framework Programme (FP/2007-2013) / ERC Grant Agreement n. 306476. This research was supported in part by the National Science Foundation under Grant No. NSF PHY11-25915. 
SS acknowledges support from Netherlands Organization for Scientific Research (NWO), VENI grant 639.041.233

%%%%%%% BIBLIOGRAPHY %%%%%%%%%%%%%%%%%%%%%%%%%%%%%%%%%%%%%%%%%%%%%%%%%%%%%%%%%%

\label{lastpage}

\end{document}